\documentclass[twocolumn,10pt,journal,a4paper]{IEEEtran}
\usepackage{bbm}
\usepackage[latin1]{inputenc}
\usepackage{amssymb}
\usepackage{latexsym}
\usepackage{mathrsfs}
\usepackage{amsfonts}
\usepackage{amsbsy}
\usepackage{amsmath}
\usepackage{times}
\usepackage[usenames]{color}
\usepackage{epsfig,epsf,amssymb,amsmath}
\newcommand{\bm}[1]{\mbox{\boldmath{$#1$}}}
\input epsf

\title{\huge Blind  Adaptive Reduced-Rank Detectors for DS-UWB Systems
Based on Joint Iterative Optimization and the Constrained Constant
Modulus Criterion}
\author{Sheng Li and Rodrigo C. de Lamare \vspace{-2em}
\thanks{This work is supported by the Department of Electronics, University of York.
The authors are with the Communications Research Group, Department of
Electronics, University of York, York, YO10 5DD, UK (e-mail: \{sl546 and
rcdl500\}@ohm.york.ac.uk). }}
\begin{document}
\maketitle

\begin{abstract}
A novel linear blind adaptive receiver based on joint iterative optimization
(JIO) and the constrained constant modulus (CCM) design criterion is proposed
for interference suppression in direct-sequence ultra-wideband (DS-UWB)
systems. The proposed blind receiver consists of two parts, a transformation
matrix that performs dimensionality reduction and a reduced-rank filter that
produces the output. In the proposed receiver, the transformation matrix and
the reduced-rank filter are updated jointly and iteratively to minimize the
constant modulus (CM) cost function subject to a constraint. Adaptive
implementations for the JIO receiver are developed by using the normalized
stochastic gradient (NSG) and recursive least-squares (RLS) algorithms. In
order to obtain a low-complexity scheme, the columns of the transformation
matrix with the RLS algorithm are updated individually. Blind channel
estimation algorithms for both versions (NSG and RLS) are implemented. Assuming
the perfect timing, the JIO receiver only requires the spreading code of the
desired user and the received data. Simulation results show that both versions
of the proposed JIO receivers have excellent performance in suppressing the
inter-symbol interference (ISI) and multiple access interference (MAI) with a
low complexity.
\end{abstract}

\section{Introduction}
\IEEEPARstart{U}{ltra}-wideband (UWB) technology
\cite{MZwin1998}-\cite{MWolf2009}, which can achieve very high data rate, is a
promising short-range wireless communication technique. By spreading the
information symbols with a pseudo-random (PR) code, direct sequence (DS)-UWB
technique enables multiuser communications \cite{oppermann2004}. In the DS-UWB
systems, a high degree of diversity is achieved at the receiver due to the
large number of resolvable multipath components (MPCs) \cite{dajana2007}.
Receivers are required to efficiently suppress the severe inter-symbol
interference (ISI) that is caused by the dense multipath channel and the
multiple-access interference (MAI) that is caused by the lack of orthogonality
between signals at the receiver in multiuser communications.

Blind adaptive linear receivers \cite{Mhongi1995}-\cite{joaquin1998} are
efficient schemes for interference suppression as they offer higher spectrum
efficiency than the adaptive schemes that require a training stage. Low
complexity blind receiver designs can be obtained by solving constrained
optimization problems based on the constrained constant modulus (CCM) or
constrained minimum variance (CMV) criterion
\cite{Rodrigo2008add},\cite{gsbiradar2008}. The blind receiver designs based on
the CCM criterion have shown better performance and increased robustness
against signature mismatch over the CMV approaches
\cite{Rodrigo2008add},\cite{joaquin1998}. Recently, blind full-rank
stochastic gradient (SG) and RLS adaptive filters based on the
constrained optimization have been proposed for multiuser detection in DS-UWB
communications \cite{gsbiradar2008},\cite{JLIU2007}. For DS-UWB systems in
which the received signal length is large due to the long channel delay spread,
the interference sensitive full-rank adaptive schemes experience slow
convergence rate. In the large filter scenarios, the reduced-rank algorithms
can be adopted to accelerate the convergence and provide an increased
robustness against interference and noise.

By projecting the received signal onto a lower-dimensional subspace
and adapting a lower-order filter to process the reduced-rank
signal, the reduced-rank filters can achieve faster convergence than
the full-rank schemes \cite{AMHaimovich1991}-\cite{Rodrigo2007}. The
existing reduced-rank schemes include the eigen-decomposition
methods and the Krylov subspace schemes. The eigen-decomposition
methods include the principal components (PC) \cite{AMHaimovich1991}
and the cross-spectral metric (CSM) \cite{JSGoldstein1997}, which
are based on the eigen-decomposition of the estimated covariance
matrix of the received signal. In the PC scheme, the received signal
is projected onto a subspace associated with the largest
eigen-values \cite{MLHonig2002} and in the CSM approach, the
subspace is selected with maximum signal to interference and noise
ratio (SINR) \cite{JohnDavidHiemstra}. It is known that the optimal
representation of the input data can be obtained by the
eigen-decomposition of its covariance matrix $\mathbf R$
\cite{JSGoldstein1998}.
However, %, $\mathbf R$ is unknown and must be
%estimated. In addition,
these methods have very high computational complexity and the robustness
against interference is often poor in heavily loaded communication systems
\cite{MLHonig2002}. The Krylov subspace schemes include the powers of R (POR)
\cite{wanshichen2002}, the multistage Wiener filter (MSWF)
\cite{MLHonig2002},\cite{JSGoldstein1998} and the auxiliary vector filtering
(AVF) \cite{dapados2001}. All these schemes project the received signal onto
the Krylov subspace \cite{wanshichen2002} and achieve faster convergence speed
than the full-rank schemes with a smaller filter size. However, the high
computational complexity is also a problem of the Krylov subspace methods.

For the UWB systems, the reduced-rank receivers that require training sequences
have recently been developed in \cite{jianzhang2004}-\cite{Sheng2010vtc}.
Solutions for reduced-rank channel estimation and synchronization in single
user UWB systems have been proposed in \cite{jianzhang2004}. For multiuser
detection in UWB communications, reduced-rank schemes have been developed in
\cite{shwu2004}-\cite{yTian2006} that require the knowledge of the multipath
channel. We proposed a low-complexity reduced-rank interference suppression
scheme for DS-UWB systems in \cite{Sheng2010vtc}, which is able to suppress
both of the ISI and MAI efficiently. In \cite{Zxu2005}, a blind subspace
multiuser detection scheme is proposed for UWB systems which requires the
eigen-decomposition of the covariance matrix of the received signal. In this
work, a novel CCM based joint iterative optimization (JIO) blind reduced-rank
receiver is proposed. A transformation matrix and a reduced-rank filter
construct the proposed receiver and they are updated jointly and iteratively to
minimize the CM cost function subject to a constraint. The proposed receiver
allows information exchange between the transformation matrix and the
reduced-rank filter. This distinguishing feature leads to a more efficient
adaptive implementation than the existing reduced-rank schemes. Note that the
constraint is necessary since it enables us to avoid the undesired local
minima. The adaptive NSG and RLS algorithms are developed for the JIO receiver.
In the NSG version, a low-complexity leakage SG channel estimator that was
proposed in \cite{xgdoukopoulos2005} is adopted. Applying an approximation to
the covariance matrix of the received signal, the RLS channel estimator
proposed in \cite{xgdoukopoulos2005} is modified for the proposed JIO-RLS with
reduced complexity. Since each column of the transformation matrix can be
considered as a direction vector on one dimension of the subspace, we update
the transformation matrix column by column to achieve a better
representation of the projection procedure in the JIO-RLS. %We remark
%that the JIO-RLS is the most practically important contribution in
%this work, which achieves better tradeoffs between the complexity
%and the performance than the JIO-NSG in DS-UWB systems.

The main contributions of this work are summarized as follows:

\begin{itemize}
\item {A novel linear blind JIO reduced-rank receiver based on the CCM criterion is proposed for interference suppression in DS-UWB systems.}
\end{itemize}

\begin{itemize}
\item {NSG algorithms, which are able to facilitate the setting of step sizes in multiuser
scenarios, are developed for the proposed reduced-rank receivers.}
\end{itemize}

\begin{itemize}
\item {RLS algorithms are developed to jointly update the columns of the transformation matrix and the reduced-rank filter with low complexity.}
\end{itemize}

\begin{itemize}
\item {A rank adaptation algorithm is developed to achieve a better tradeoff between the convergence speed and the steady state performance.}
\end{itemize}

\begin{itemize}
\item {The convergence properties of the CM cost function with a constraint are discussed.}
\end{itemize}

\begin{itemize}
\item {Simulations are performed with the IEEE 802.15.4a channel models and severe ISI and MAI are assumed for the evaluation of the proposed scheme
against existing techniques.}
\end{itemize}

The rest of this paper is structured as follows. Section \ref{sec:uwbsystem}
presents the DS-UWB system model. The design of the JIO CCM blind receiver is
detailed in Section \ref{sec:receiverdesign}. The proposed NSG and RLS versions
of the blind JIO receiver are described in Section \ref{sec:proposedjionsg} and
\ref{sec:proposedjioca}, respectively. In Section \ref{sec:complexityanalysis},
a complexity analysis for the proposed receiver versions is detailed and a rank
adaptation algorithm is developed for the JIO receiver. Simulation results are
shown in Section \ref{sec:simulations} and conclusions are drawn in Section
\ref{sec:conclusion}.

%%%%%%%%%%%%%%%%%%%%%%%%%%%%%%%%%%%%%%%%%%%%%%%%%%%%%%%%%%%%%%%%%%%%%%%%%%%%%%%%%%%%%%%%%%%%%%%%%%%%%%%%%%%%%%%%%%%
%%%%%%%%%%%%%%%%%%%%%%%%%%%%%%%%%%%%%%%%%%%%%%%%%%%%%%%%%%%%%%%%%%%%%%%%%%%%%%%%%%%%%%%%%%%%%%%%%%%%%%%%%%%%%%%%%%%
\section{DS-UWB System Model}
\label{sec:uwbsystem}

In this work, we consider the uplink of a binary phase-shift keying (BPSK)
DS-UWB system with $K$ users. A random spreading code $\mathbf {s}_{k}$ is
assigned to the $k$-th user with a spreading gain $N_{c} = T_{s}/T_{c}$, where
$T_{s}$ and $T_{c}$ denote the symbol duration and chip duration, respectively.
The transmit signal of the $k$-th user (where $k=1,2,\dots,K$) can be expressed
as
%%%%%%%%%%%%%%%%%%%%%%%%%%%%%%%%%%%%%%%%%%%%%%%%%%%%%%%%%%%%%%%%%%%%%%%%%%%%%%%%%%%%%%%%%%%%%%%%%%%%%%%%
\begin{equation}
x^{(k)}(t)=\sqrt{E_{k}}\sum_{i=-{\infty}}^{\infty}\sum_{j=0}^{N_{c}-1}
p_{t}(t-iT_{s}-jT_{c})s_{k}(j)b_{k}(i),
\end{equation}where $b_{k}(i)$ $\in$ $\{\pm1\}$
denotes the BPSK symbol for the $k$-th user at the $i$-th time
instant, $s_{k}(j)$ denotes the $j$-th chip of the spreading code
$\mathbf {s}_{k}$ (where $j=1,2,\dots,N_{c}$). $E_{k}$ denotes the
transmission energy of the $k$-th user. $p_{t}(t)$ is the pulse
waveform of width $T_{c}$. Throughout this paper, the pulse waveform
$p_{t}(t)$ is modeled as the root-raised cosine (RRC) pulse with a
roll-off factor of $0.5$ \cite{RFisher2005},\cite{Aparihar2007}. {
The channel model considered is the IEEE 802.15.4a channel model for
the indoor residential environment \cite{Molisch2005}.} This
standard channel model includes some generalizations of the
Saleh-Valenzuela model and takes the frequency dependence of the
path gain into account \cite{Molisch2006}. In addition, the 15.4a
channel model is valid for both low-data-rate and high-data-rate UWB
systems \cite{Molisch2006}. For the $k$-th users, the channel
impulse response (CIR) of the standard channel model can be
expressed as
%%%%%%%%%%%%%%%%%%%%%%%%%%%%%%%%%%%%%%%%%%%%%%%%%%%%%%%%%%%%%%%%%%%%%%%%%%%%%%%%%%%%%%%%%%%%%%%%%%%%%%%%
\begin{equation}
h_{k}(t)=
\sum_{u=0}^{L_{c}-1}\sum_{v=0}^{L_{r}-1}\alpha_{u,v}e^{j\phi_{u,v}}\delta(t-T_{u}-T_{u,v}),
\end{equation} where $L_{c}$ denotes the number of clusters, $L_{r}$ is the
number of MPCs in one cluster. $\alpha_{u,v}$ is the fading gain of the $v$-th
MPC in the $u$-th cluster, $\phi_{u,v}$ is uniformly distributed in $[0,2\pi)$.
$T_{u}$ is the arrival time of the $u$-th cluster and $T_{u,v}$ denotes the
arrival time of the $v$-th MPC in the $u$-th cluster. For the sake of
simplicity, we express the CIR as
\begin{equation}
h_{k}(t)= \sum_{l=0}^{L-1}h_{k,l}\delta(t-lT_{\tau}),\label{eq:channel}
\end{equation} where $h_{k,l}$
and $lT_{\tau}$ present the complex-valued fading factor and the arrival time
of the $l$-th MPC ($l=uL_{c}+v$), respectively. $L=T_{DS}/T_{\tau}$ denotes the
total number of MPCs where $T_{DS}$ is the channel delay spread. Assuming that
the timing is acquired, the received signal can be expressed as
%%%%%%%%%%%%%%%%%%%%%%%%%%%%%%%%%%%%%%%%%%%%%%%%%%%%%%%%%%%%%%%%%%%%%%%%%%%%%%%%%%%%%%%%%%%%%%%%%%%%%%%%
\begin{equation}
z(t)= \sum_{k=1}^{K}\sum_{l=0}^{L-1} h_{k,l}x^{(k)}(t-lT_{\tau})+n(t),
\end{equation} where $n(t)$ is the additive white Gaussian noise (AWGN) with
zero mean and a variance of $\sigma_{n}^{2}$. This signal is first passed
through a chip-matched filter (MF) and then sampled at the chip rate. We select
a total number of $M=(T_{s}+T_{DS})/T_{c}$ observation samples for the
detection of each data bit, where $T_{s}$ is the symbol duration, $T_{DS}$ is
the channel delay spread and $T_{c}$ is the chip duration. Assuming the
sampling starts at the zero-th time instant, then the $m$-th sample is given by
%%%%%%%%%%%%%%%%%%%%%%%%%%%%%%%%%%%%%%%%%%%%%%%%%%%%%%%%%%%%%%%%%%%%%%%%%%%%%%%%%%%%%%%%%%%%%%%%%%%%%%%%
\begin{equation*}
r_{m}= \int_{mT_{c}}^{(m+1)T_{c}} z(t)p_{r}(t)~ dt, ~~m=1,2,\dots,M
\end{equation*} where $p_{r}(t)=p_{t}^{*}(-t)$ denotes the chip-matched filter, $(\cdot)^{*}$ denotes
the complex conjugation. After the chip-rate sampling, the discrete-time
received signal for the $i$-th data bit can be expressed as $\mathbf
r(i)=[r_{1}(i),r_{2}(i),\dots,r_{M}(i)]^{T}$, where $(\cdot)^T$ is the
transposition and we can further express it in a matrix form as
%%%%%%%%%%%%%%%%%%%%%%%%%%%%%%%%%%%%%%%%%%%%%%%%%%%%%%%%%%%%%%%%%%%%%%%%%%%%%%%%%%%%%%%%%%%%%%%%%%%%%%%%
\begin{equation}
%\begin{split}
\mathbf r(i)=\sum_{k=1}^{K}\sqrt{E_{k}}\mathbf P_{r}\mathbf
{H}_{k}\mathbf P_{t}\mathbf s_{k}
b_{k}(i)+\mbox{\boldmath$\eta$}(i)+\mathbf
{n}(i),%\\
%&=\sum_{k=1}^{K}\mathbf P_{r}\mathbf {S}_{e,k}\mathbf
%h_{k}b_{k}(i)+\mbox{\boldmath$\eta$}(i)+\mathbf
%{n}(i),\\
%\end{split}
\end{equation} where $\mathbf {H}_{k}$ is the Toeplitz channel matrix for the
$k$-th user with the first column being the CIR $\mathbf
h_{k}=[h_{k}(0),h_{k}(1),\dots,h_{k}(L-1)]^{T}$ zero-padded to length
$M_{H}=(T_{s}/T_{\tau})+L-1$. The matrix $\mathbf P_{r}$ represents the MF and
chip-rate sampling with the size $M$-by-$M_{H}$. $\mathbf P_{t}$ denotes the
$(T_{s}/T_{\tau})$-by-$N_{c}$ pulse shaping matrix. In order to facilitate the
blind channel estimation in a later development, we rearrange the term and
express the received signal as
%%%%%%%%%%%%%%%%%%%%%%%%%%%%%%%%%%%%%%%%%%%%%%%%%%%%%%%%%%%%%%%%%%%%%%%%%%%%%%%%%%%%%%%%%%%%%%%%%%%%%%%%
\begin{equation}
\mathbf r(i) =\sum_{k=1}^{K}\sqrt{E_{k}}\mathbf P_{r}\mathbf
{S}_{e,k}\mathbf h_{k}b_{k}(i)+\mbox{\boldmath$\eta$}(i)+\mathbf
{n}(i),
\end{equation}where $\mathbf {S}_{e,k}$ is the
Toeplitz matrix with the first column being the vector $\mathbf
s_{e,k}=\mathbf P_{t}\mathbf s_{k}$ zero-padded to length $M_{H}$.
The vector $\mbox{\boldmath$\eta$}(i)$ denotes the ISI from $2G$
adjacent symbols, where $G$ denotes the minimum integer that is
larger than or equal to the scalar term $T_{DS}/T_{s}$. Here, we
express the ISI vector in a general form that is given by
%%%%%%%%%%%%%%%%%%%%%%%%%%%%%%%%%%%%%%%%%%%%%%%%%%%%%%%%%%%%%%%%%%%%%%%%%%%%%%%%%%%%%%%%%%%%%%%%%%%%%%%%
\begin{equation}
\begin{split}
\mbox{\boldmath$\eta$}(i)&=\sum_{k=1}^{K}\sum_{g=1}^{G}\sqrt{E_{k}}\mathbf
P_{r}\mathbf {H}_{k}^{(-g)}\mathbf P_{t}\mathbf s_{k}
b_{k}(i-g)\\
&+\sum_{k=1}^{K}\sum_{g=1}^{G}\sqrt{E_{k}}\mathbf P_{r}\mathbf
{H}_{k}^{(+g)}\mathbf P_{t}\mathbf s_{k} b_{k}(i+g),
\end{split}
\end{equation} where the channel matrices for the ISI
are given by
%%%%%%%%%%%%%%%%%%%%%%%%%%%%%%%%%%%%%%%%%%%%%%%%%%%%%%%%%%%%%%%%%%%%%%%%%%%%%%%%%%%%%%%%%%%%%%%%%%%%%%%%
\begin{equation}
\mathbf {H}_{k}^{(-g)}=\begin{bmatrix}\mathbf 0 &\mathbf H_{k}^{(u,g)}\\
\mathbf 0& \mathbf 0
\end{bmatrix}~~;~~\mathbf {H}_{k}^{(+g)}=\begin{bmatrix}\mathbf 0 &\mathbf 0\\ \mathbf H_{k}^{(l,g)} & \mathbf 0
\end{bmatrix}.
\end{equation} Note that the matrices $\mathbf H_{k}^{(u,g)}$ and $\mathbf H_{k}^{(l,g)}$ have the same size
as $\mathbf {H}_{k}$, which is $M_{H}$-by-$(T_{s}/T_{\tau})$, and
can be considered as the partitions of an upper triangular matrix
$\mathbf H_{\rm up}$ and a lower triangular matrix $\mathbf H_{\rm
low}$, respectively, where
%%%%%%%%%%%%%%%%%%%%%%%%%%%%%%%%%%%%%%%%%%%%%%%%%%%%%%%%%%%%%%%%%%%%%%%%%%%%%%%%%%%%%%%%%%%%%%%%%%%%%%%%
\begin{equation*}
\begin{split}
&\mathbf H_{\rm up}=\begin{bmatrix}h_{k}(L-1)     & \dots & h_{k}(L-\frac{T_{DS}-(g-1)T_{s}}{T_{\tau}})\\
                                         & \ddots       &\vdots\\
                                         &        &h_{k}(L-1)
\end{bmatrix}~;\\
&\mathbf H_{\rm low}=\begin{bmatrix}h_{k}(0) & & \\
                                               \vdots  &\ddots   &\\
                                               h_{k}(\frac{T_{DS}-(g-1)T_{s}}{T_{\tau}}-2)&\dots
                                               &h_{k}(0)
\end{bmatrix}.
\end{split}
\end{equation*} These triangular matrices have the row-dimension of
$[T_{DS}-(g-1)T_{s}]/T_{\tau}-1=L-(g-1)T_{s}/T_{\tau}-1$. Note that
when the channel delay spread is large, the row-dimension of these
triangular matrices could surpass the column dimension of the matrix
$\mathbf {H}_{k}$, which is $T_{s}/T_{\tau}$. Hence, in case of
\begin{equation}
\begin{split}
&L-(g-1)T_{s}/T_{\tau}-1>T_{s}/T_{\tau}, \\
i.e. ~~~&L>gT_{s}/T_{\tau}+1,
\end{split}
\end{equation} the matrix $\mathbf H_{k}^{(u,g)}$ is the last $T_{s}/T_{\tau}$ columns of the
upper triangular matrix $\mathbf H_{\rm up}$ and $\mathbf H_{k}^{(l,g)}$ is the
first $T_{s}/T_{\tau}$ columns of the lower triangular matrix $\mathbf H_{\rm
low}$. When $L<gT_{s}/T_{\tau}+1$, $\mathbf H_{k}^{(u,g)}=\mathbf H_{\rm up}$
and $\mathbf H_{k}^{(l,g)}=\mathbf H_{\rm low}$. It is interesting to review
the expression of the ISI vector via its physical meaning, since the
row-dimension of the matrices $\mathbf H_{k}^{(u,g)}$ and $\mathbf
H_{k}^{(l,g)}$, which is $L-(g-1)T_{s}/T_{\tau}-1$, reflects the time domain
overlap between the data symbol $b(i)$ and the adjacent symbols of $b(i-g)$ and
$b(i+g)$.

%%%%%%%%%%%%%%%%%%%%%%%%%%%%%%%%%%%%%%%%%%%%%%%%%%%%%%%%%%%%%%%%%%%%%%%%%%%%%%%%%%%%%%%%%%%%%%%%%%%%%%%%%%%%%%%%%%%%
%%%%%%%%%%%%%%%%%%%%%%%%%%%%%%%%%%%%%%%%%%%%%%%%%%%%%%%%%%%%%%%%%%%%%%%%%%%%%%%%%%%%%%%%%%%%%%%%%%%%%%%%%%%%%%%%%%%%
\section{Proposed Blind JIO Reduced-rank Receiver Design}
\label{sec:receiverdesign}
\begin{figure}[htb]
\begin{minipage}[h]{1.0\linewidth}
  \centering
  \centerline{\epsfig{figure=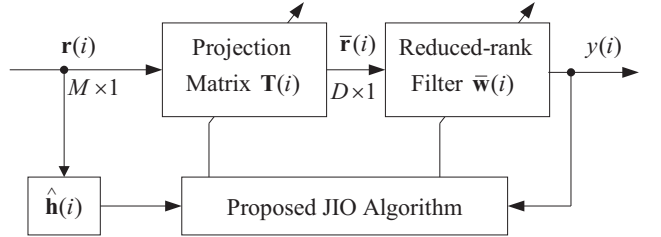,scale=0.75}}
\end{minipage}
\caption{Block diagram of the proposed blind reduced-rank receiver.}
\label{fig:blockdiagram}
\end{figure}
%%%%%%%%%%%%%%%%%%%%%%%%%%%%%%%%%%%%%%%%%%%%%%%%%%%%%%%%%%%%%%%%%%%%%
%%%%%%%%%%%%%%%%%%%%%%%%%%%%%%%%%%%%%%%%%%%%%%%%%%%%%%%%%%%%%%%%%%%%%
In this section, we detail the design of the proposed JIO reduced-rank receiver
that is able to recover the data symbol from the noisy received signal blindly.
The block diagram of the proposed receiver is shown in
Fig.\ref{fig:blockdiagram}. In the JIO blind linear receiver, the reduced-rank
received signal can be expressed as
\begin{equation}
\mathbf {\bar r}(i)=\mathbf T^{H}(i)\mathbf r(i),
\end{equation} where
$\mathbf T(i)$ is the $M$-by-$D$ (where $D\ll M$) transformation matrix. After
the projection, $\mathbf {\bar r}(i)$ is fed into the reduced-rank filter
$\bar{\mathbf{w}}(i)$ and the output signal is given by
\begin{equation}
y(i)=\bar{\mathbf{w}}^{H}(i)\mathbf {\bar r}(i).
\end{equation} The decision of the desired data symbol is defined as
\begin{equation}
\hat {b}(i)={\rm sign}(\mathfrak{R}[y(i)]).
\end{equation}where ${\rm sign}(\cdot)$ is the algebraic sign function and
$\mathfrak{R}(\cdot)$ represents the real part of a complex number.

The optimization problem to be solved can be expressed as
%%%%%%%%%%%%%%%%%%%%%%%%%%%%%%%%%%%%%%%%%%%%%%%%%%%%%%%%%%%%%%%%%%%%%%%%%%%%%%%%%%%%%%%%%%%%%%%%%%%%%%%%
\begin{equation}
\left[{\mathbf{\bar w}}(i),\mathbf{T}(i)\right]\\
=\mathop{\mbox{arg}} \mathop{\mbox{min}}_{{\mathbf{\bar w}}(i),\mathbf{T}(i)}
\mathbf J_{\rm JIO} \big(\bar{\mathbf{w}}(i),\mathbf{T}(i)\big),
\end{equation} subject to the constraint
\begin{equation}
\bar{\mathbf w} ^{H}(i) \mathbf T^{H}(i) \mathbf p=v,
\end{equation} where $\mathbf p=\mathbf P_{r}\mathbf {S}_{e}\mathbf h$ is
defined as the effective signature vector for the desired user and $v$ is a
real-valued constant to ensure the convexity of the CM cost function
%%%%%%%%%%%%%%%%%%%%%%%%%%%%%%%%%%%%%%%%%%%%%%%%%%%%%%%%%%%%%%%%%%%%%
\begin{equation}
\mathbf J_{\rm JIO}
\big(\bar{\mathbf{w}}(i),\mathbf{T}(i)\big)=\frac{1}{2}E\left[(\left|y(i)\right|^{2}-1)^{2}\right].\label{eq:costfuncjirrlms}
\end{equation} The convergence
properties of the CM cost function subject to a constraint are discussed in
Appendix \ref{app:convergence}.

Let us now consider the problem through the Lagrangian
%%%%%%%%%%%%%%%%%%%%%%%%%%%%%%%%%%%%%%%%%%%%%%%%%%%%%%%%%%%%%%%%%%%%%
\begin{equation}
\mathcal {L}_{\rm JIO}
\big(\bar{\mathbf{w}}(i),\mathbf{T}(i)\big)=\frac{1}{2}E\left[(\left|y(i)\right|^{2}-1)^{2}\right]+\mathfrak{R}[\lambda(i)(\bar{\mathbf
w} ^{H}(i) \mathbf T^{H}(i) {\mathbf p}-v)],\label{eq:costfuncjirrlms22}
\end{equation}where $\lambda(i)$ is a complex-valued Lagrange multiplier.
In order to obtain the adaptation equation of $\mathbf T(i)$, we firstly assume
that $\bar{\mathbf{w}}(i)$ is fixed and the gradient of the Lagrangian with
respect to $\mathbf T(i)$ is given by
\begin{equation}
\nabla_{T}\mathcal {L}_{\rm JIO}=E\left[e(i)y^{*}(i)\mathbf r(i)\bar{\mathbf
w}^{H}(i)\right]+\frac{\lambda_{T}(i)}{2}{\mathbf p}\bar{\mathbf w}^{H}(i),
\label{eq:gradientoptt}
\end{equation} where $\lambda_{T}(i)$ is the complex-valued Lagrange
multiplier for updating the transformation matrix and $e(i)=|y(i)|^{2}-1$ is
defined as a real-valued error signal. Recalling the relationship
$y^{*}(i)=\mathbf r^{H}(i)\mathbf T(i)\bar{\mathbf{w}}(i)$ and setting
\eqref{eq:gradientoptt} to a zero matrix, we obtain
%%%%%%%%%%%%%%%%%%%%%%%%%%%%%%%%%%%%%%%%%%%%%%%%%%%%%%%%%%%%%%%%%%%%%%%%%%%%%%%%%%
\begin{equation}
\mathbf T_{\rm opt}=\mathbf R_{Y}^{-1}\left(\mathbf
D_{T}-\frac{\lambda_{T}(i)}{2}\mathbf p\mathbf {\bar
w}^{H}(i)\right)\mathbf R_{w}^{-1},
\end{equation} where $\mathbf R_{Y}=E[|y(i)|^{2}\mathbf r(i)\mathbf
r^{H}(i)]$, $\mathbf D_{T}=E[y^{*}(i)\mathbf r(i)\mathbf {\bar
w}^{H}(i)]$ and $\mathbf R_{w}=E[\mathbf {\bar w}(i)\mathbf {\bar
w}^{H}(i)]$. Using the constraint $\bar{\mathbf w} ^{H}(i) \mathbf
T^{H}_{\rm opt} \mathbf p=v$, we obtain the Lagrange multiplier
%%%%%%%%%%%%%%%%%%%%%%%%%%%%%%%%%%%%%%%%%%%%%%%%%%%%%%%%%%%%%%%%%%%%%%%%%%%%%%%%%%
\begin{equation}
\lambda_{T}(i)=2\left(\frac{\mathbf {\bar w}^{H}(i)\mathbf
R_{w}^{-1}\mathbf D_{T}\mathbf R_{Y}^{-1}\mathbf p-v}{\mathbf
{\bar w}^{H}(i)\mathbf R_{w}^{-1}\mathbf {\bar w}(i)\mathbf
p^{H}\mathbf R_{Y}^{-1}\mathbf p}\right)^{*}.
\end{equation}

Now, we assume that $\mathbf T(i)$ is fixed in
\eqref{eq:costfuncjirrlms22} and calculate the gradient of the
Lagrangian with respect to $\mathbf {\bar w}(i)$, which is given by
\begin{equation}
\nabla_{w}\mathcal {L}_{\rm JIO}=E\left[e(i)\mathbf T^{H}(i)\mathbf
r(i)y^{*}(i)\right]+\frac{\lambda_{w}(i)}{2}\mathbf T^{H}(i){\mathbf
p}, \label{eq:gradientopt}
\end{equation} where $\lambda_{w}(i)$ is the complex-valued Lagrange
multiplier for updating the reduced-rank filter. Rearranging the
terms, we obtain
%%%%%%%%%%%%%%%%%%%%%%%%%%%%%%%%%%%%%%%%%%%%%%%%%%%%%%%%%%%%%%%%%%%%%%%%%%%%%%%%%%
\begin{equation}
\mathbf {\bar w}_{\rm opt}=\mathbf R_{\bar y}^{-1}\left(\mathbf
d_{\bar r}-\frac{\lambda_{w}(i)}{2}\mathbf T^{H}(i)\mathbf p\right),
\end{equation} where $\mathbf R_{\bar y}=E[|y(i)|^{2}\mathbf {\bar r}(i)\mathbf {\bar r}^{H}(i)]$
and $\mathbf d_{\bar r}=E[y^{*}(i)\mathbf {\bar r}(i)]$. Using the
constraint $\bar{\mathbf w}_{\rm opt} ^{H}\mathbf T^{H}(i) \mathbf
p=v$, we obtain the Lagrange multiplier
%%%%%%%%%%%%%%%%%%%%%%%%%%%%%%%%%%%%%%%%%%%%%%%%%%%%%%%%%%%%%%%%%%%%%%%%%%%%%%%%%%
\begin{equation}
\lambda_{w}(i)=2\left(\frac{\mathbf d_{\bar r}^{H}\mathbf R_{\bar
y}^{-1}\mathbf T^{H}(i)\mathbf p-v}{\mathbf p^{H}\mathbf
T(i)\mathbf R_{\bar y}^{-1}\mathbf T^{H}(i)\mathbf p}\right)^{*}.
\end{equation}

With the solutions of $\mathbf T_{\rm opt}$ and $\mathbf {\bar w}_{\rm opt}$,
the NSG and RLS adaptive versions of the JIO receiver will be developed in the
following sections, in which the direct matrix inversions are not required and
the computational complexity is reduced. Note that when adaptive algorithms are
implemented to estimate $\mathbf T_{\rm opt}$ and $\mathbf {\bar w}_{\rm opt}$,
$\mathbf T(i)$ is a function of $\mathbf {\bar w}(i)$ and $\mathbf {\bar w}(i)$
is a function of $\mathbf T(i)$. Thus, the optimal CCM design is not in a
closed form and one possible solution for such optimization problem is to
jointly and iteratively adapt these two quantities. The joint update means for
the $i$-th time instant, $\mathbf T(i)$ is updated with the knowledge of
$\mathbf T(i-1)$ and $\mathbf {\bar w}(i-1)$, then $\mathbf {\bar w}(i)$ is
updated with $\mathbf T(i)$ and $\mathbf {\bar w}(i-1)$. Each iterative update
can be considered as one repetition of the joint update.

It should also be noted that the blind JIO receiver design requires
the knowledge of the effective signature vector of the desired user,
or equivalently, the channel parameters. In this work, the channel
coefficients are not given and must be estimated. Here, we employ
the variant of the power method introduced in
\cite{xgdoukopoulos2005}:
%%%%%%%%%%%%%%%%%%%%%%%%%%%%%%%%%%%%%%%%%%%%%%%%%%%%%%%%%%%%%%%%%%%%%
\begin{equation}
\mathbf {\hat h}(i)=\left(\mathbf I-\mathbf {\hat V}(i)/tr[\mathbf
{\hat V}(i)]\right)\mathbf {\hat h}(i-1),\label{eq:hatHfull}
\end{equation} where the $L$-by-$L$ matrix is defined as
%%%%%%%%%%%%%%%%%%%%%%%%%%%%%%%%%%%%%%%%%%%%%%%%%%%%%%%%%%%%%%%%%%%%%
\begin{equation}
\mathbf {\hat V}(i)=\mathbf {S}_{e}^{H}\mathbf P_{r}^{H}\mathbf
R^{-m}(i)\mathbf P_{r}\mathbf {S}_{e},\label{eq:hatV}
\end{equation}
and $\mathbf I$ is the identity matrix, $tr[\cdot]$ stands for trace and we
make $\mathbf {\hat h}(i)\leftarrow\mathbf {\hat h}(i)/\|\mathbf {\hat h}(i)\|$
to normalize the channel. $\mathbf R(i)=\sum^{i}_{j=1}\alpha^{i-j}\mathbf
r(j)\mathbf r^{H}(j)$ and $m$ is a finite power. The estimate of the matrix
$\mathbf R^{-1}(i)$ is obtained recursively via the matrix inversion lemma
\cite{haykin} and is given by
%%%%%%%%%%%%%%%%%%%%%%%%%%%%%%%%%%%%%%%%%%%%%%%%%%%%%%%%%%%%%%%%%%%%%
\begin{equation}
\mathbf {\hat R}^{-1}(i)=\frac{1}{\alpha}\left(\mathbf {\hat
R}^{-1}(i-1)-(\phi(i)\bm {\kappa}(i))\bm {\kappa}^{H}(i) \right),
\label{eq:CArlsinvR}
\end{equation}where $\alpha$ is the forgetting factor, $\bm {\kappa} (i)=\mathbf {\hat R}^{-1}(i-1)\mathbf
r(i)$ and $\phi(i)=\left(\alpha + \mathbf r^{H}(i)\bm {\kappa}
(i)\right)^{-1}$. The estimation of the inversion of the covariance matrix
requires $3M^{2}+2{M}+1$ multiplications and $2M^{2}$ additions. Equation
\eqref{eq:hatV} requires $(m+1)M^{2}L$ multiplications and
$(m+1)M^{2}L-(m+1)ML$ additions, while equation \eqref{eq:hatHfull} requires
$L^{2}$ multiplications and $L^{2}+L-1$ additions (the multiplications and
additions in this work are both complex-valued operations). Note that, the
matrix $\mathbf P_{r}\mathbf {S}_{e}$ is assumed given at the receiver.

The estimate of the effective signature vector can be finally
obtained as
%%%%%%%%%%%%%%%%%%%%%%%%%%%%%%%%%%%%%%%%%%%%%%%%%%%%%%%%%%%%%%%%%%%%%
\begin{equation}
\mathbf {\hat p}(i)=\mathbf P_{r}\mathbf {S}_{e}\mathbf {\hat h}(i),
\end{equation} where $\mathbf {\hat h}(i)$ is given in
\eqref{eq:hatHfull}.

%%%%%%%%%%%%%%%%%%%%%%%%%%%%%%%%%%%%%%%%%%%%%%%%%%%%%%%%%%%%%%%%%%%%%%%%%%%%%%%%%%%%%%%%%%%%%%%%%%%%%%%%%%%%%%%%%%%
%%%%%%%%%%%%%%%%%%%%%%%%%%%%%%%%%%%%%%%%%%%%%%%%%%%%%%%%%%%%%%%%%%%%%%%%%%%%%%%%%%%%%%%%%%%%%%%%%%%%%%%%%%%%%%%%%%%
\section{Proposed JIO-NSG Algorithms}
\label{sec:proposedjionsg}
%%%%%%%%%%%%%%%%%%%%%%%%%%%%%%%%%%%%%%%%%%%%%%%%%%%%%%%%%%%%%%%%%%%%%%
%\begin{figure}[htb]
%\begin{minipage}[h]{1.0\linewidth}
%  \centering
%  \centerline{\epsfig{figure=JIOCASCHEME.eps,scale=0.75}}
%\end{minipage}
%\caption{Block diagram of the proposed blind reduced-rank receiver.}
%\label{fig:blockdiagram}
%\end{figure}
%%%%%%%%%%%%%%%%%%%%%%%%%%%%%%%%%%%%%%%%%%%%%%%%%%%%%%%%%%%%%%%%%%%%%%
%%%%%%%%%%%%%%%%%%%%%%%%%%%%%%%%%%%%%%%%%%%%%%%%%%%%%%%%%%%%%%%%%%%%%%
%The block diagram of the proposed blind JIO reduced-rank receiver is shown in
%Fig.\ref{fig:blockdiagram}. In the JIO blind linear receiver, the reduced-rank
%received signal can be expressed as $\mathbf {\bar r}(i)=\mathbf
%T^{H}(i)\mathbf r(i)$, where $\mathbf T(i)$ is the $M$-by-$D$ (where $D\ll M$)
%transformation matrix. After the projection, $\mathbf {\bar r}(i)$ is fed into the
%reduced-rank filter $\bar{\mathbf{w}}(i)$ and the output signal is given by
%$y(i)=\bar{\mathbf{w}}^{H}(i)\mathbf {\bar r}(i).$ The estimate of the desired
%data symbol is defined as $\hat {b}(i)={\rm sign}(\mathfrak{R}[y(i)])$.

In this section, we develop the NSG algorithm to jointly and
iteratively update $\mathbf T(i)$ and $\bar{\mathbf{w}}(i)$. The
blind channel estimator based on the leakage SG algorithm that is
proposed in \cite{xgdoukopoulos2005} is implemented to provide the
channel coefficients.

%%%%%%%%%%%%%%%%%%%%%%%%%%%%%%%%%%%%%%%%%%%%%%%%%%%%%%%%%%%%%%%%%%%%%%%%%%%%%%%%%%%%%%%%%%%%%%%%%%%%%%%%%%%%%%%%%%%
%%%%%%%%%%%%%%%%%%%%%%%%%%%%%%%%%%%%%%%%%%%%%%%%%%%%%%%%%%%%%%%%%%%%%%%%%%%%%%%%%%%%%%%%%%%%%%%%%%%%%%%%%%%%%%%%%%%
\subsection{JIO-NSG Algorithms}
\label{sec:jionsg}

The optimization problem to be solved in the NSG version is given by
%%%%%%%%%%%%%%%%%%%%%%%%%%%%%%%%%%%%%%%%%%%%%%%%%%%%%%%%%%%%%%%%%%%%%%%%%%%%%%%%%%%%%%%%%%%%%%%%%%%%%%%%
\begin{equation}
\left[{\mathbf{\bar w}}(i),\mathbf{T}(i)\right]\\
=\mathop{\mbox{arg}} \mathop{\mbox{min}}_{{\mathbf{\bar w}}(i),\mathbf{T}(i)}
\mathbf J_{\rm JIO} \big(\bar{\mathbf{w}}(i),\mathbf{T}(i)\big),
\end{equation} subject to $\bar{\mathbf
w} ^{H}(i) \mathbf T^{H}(i) \hat{\mathbf p}(i)=v$, where
$\hat{\mathbf p}(i)$ is the estimated signature vector obtained via
blind channel estimation that will be detailed in Section
\ref{sec:bceforjionsg} and $v$ is a real-valued constant to
ensure the convexity of the cost function
%%%%%%%%%%%%%%%%%%%%%%%%%%%%%%%%%%%%%%%%%%%%%%%%%%%%%%%%%%%%%%%%%%%%%
\begin{equation}
\mathbf J_{\rm JIO-NSG}
\big(\bar{\mathbf{w}}(i),\mathbf{T}(i)\big)=\frac{1}{2}E\left[(\left|y(i)\right|^{2}-1)^{2}\right].\label{eq:costfuncjirrlms}
\end{equation}
Here, we consider the problem through the Lagrangian
%%%%%%%%%%%%%%%%%%%%%%%%%%%%%%%%%%%%%%%%%%%%%%%%%%%%%%%%%%%%%%%%%%%%%
\begin{equation}
\mathcal {L}_{\rm JIO-NSG}
\big(\bar{\mathbf{w}}(i),\mathbf{T}(i)\big)=\frac{1}{2}E\left[(\left|y(i)\right|^{2}-1)^{2}\right]+\mathfrak{R}[\lambda_{N}(i)(\bar{\mathbf
w} ^{H}(i) \mathbf T^{H}(i) \hat{\mathbf
p}(i)-v)],\label{eq:costfuncjirrlms1}
\end{equation}where $\lambda_{N}(i)$ is a complex-valued Lagrange multiplier.
For each time instant, we firstly update $\mathbf T(i)$ while assuming that
$\bar{\mathbf{w}}(i)$ is fixed. Then we adapt $\bar{\mathbf{w}}(i)$ with the
updated $\mathbf T(i)$.

The gradient of the Lagrangian with respect to $\mathbf T(i)$ is given by
\begin{equation*}
\nabla_{T}\mathcal {L}_{\rm JIO-NSG}=E\left[e(i)y^{*}(i)\mathbf
r(i)\bar{\mathbf w}^{H}(i)\right]+\frac{1}{2}\lambda_{NT}(i)\hat{\mathbf
p}(i)\bar{\mathbf w}^{H}(i),
\end{equation*} where $\lambda_{NT}(i)$ is the
complex-valued Lagrange multiplier for updating the transformation matrix and
$e(i)=|y(i)|^{2}-1$ is defined as a real-valued error signal. Using the
instantaneous estimator to the gradient vector, the SG update equation is given
by
%%%%%%%%%%%%%%%%%%%%%%%%%%%%%%%%%%%%%%%%%%%%%%%%%%%%%%%%%%%%%%%%%%%%%
\begin{equation}
\mathbf{T}(i+1)=\mathbf{T}(i)-\mu_{T}\left(e(i)y^{*}(i)\mathbf
r(i)+\frac{\lambda_{NT}(i)}{2} \hat{\mathbf p}(i)\right)\bar{\mathbf
w}^{H}(i),\label{eq:rrnlmsti1}
\end{equation} where $\mu_{T}$ is the step size for the SG algorithm
that updates the transformation matrix. Using the constraint of $\bar{\mathbf
w}^{H}(i) \mathbf T^{H}(i+1) \hat{\mathbf p}(i)=v$, we obtain that
%%%%%%%%%%%%%%%%%%%%%%%%%%%%%%%%%%%%%%%%%%%%%%%%%%%%%%%%%%%%%%%%%%%%%
\begin{equation}
\lambda_{NT}(i)=2\frac{\hat{\mathbf p}^{H}(i)\mathbf T(i)\bar{\mathbf
w}(i)-\mu_{T} e(i)y^{*}(i)\|\bar{\mathbf w}(i)\|^{2}\hat{\mathbf
p}^{H}(i)\mathbf r(i)-v}{\mu_{T}\|\bar{\mathbf w}(i)\|^{2}\|\hat{\mathbf
p}(i)\|^{2}}. \label{eq:rrnlmslambdat}
\end{equation}
%%%%%%%%%%%%%%%%%%%%%%%%%%%%%%%%%%%%%%%%%%%%%%%%%%%%%%%%%%%%%%%%%%%%%
The NSG algorithm aims at minimizing the cost function
%%%%%%%%%%%%%%%%%%%%%%%%%%%%%%%%%%%%%%%%%%%%%%%%%%%%%%%%%%%%%%%%%%%%%
\begin{equation}
\mathbf J_{\rm JIO-NSG} (\mu_{T})=\frac{1}{2}\left[\left|\bar{\mathbf w}
^{H}(i) \mathbf T^{H}(i+1) \mathbf
r(i)\right|^{2}-1\right]^{2}.\label{eq:costfuncjirrlms11}
\end{equation}Substituting \eqref{eq:rrnlmsti1} and \eqref{eq:rrnlmslambdat}
into \eqref{eq:costfuncjirrlms11} and setting the gradient vector of
\eqref{eq:costfuncjirrlms11} with respect to $\mu_{T}$ to zeros, we obtain the
solutions
%%%%%%%%%%%%%%%%%%%%%%%%%%%%%%%%%%%%%%%%%%%%%%%%%%%%%%%%%%%%%%%%%%%%%
\begin{equation*}
\mu_{T,1}=\frac{|y(i)|-1}{|y(i)|e(i)A_{T,1}},~ \mu_{T,2}=\frac{|y(i)|+
1}{|y(i)|e(i)A_{T,1}},
\end{equation*}
%%%%%%%%%%%%%%%%%%%%%%%%%%%%%%%%%%%%%%%%%%%%%%%%%%%%%%%%%%%%%%%%%%%%%
\begin{equation*}
\mu_{T,3}=\mu_{T,4}=\frac{1}{e(i)A_{T,1}},
\end{equation*}where the real-valued scale term $A_{T,1}$ is defined
as
%%%%%%%%%%%%%%%%%%%%%%%%%%%%%%%%%%%%%%%%%%%%%%%%%%%%%%%%%%%%%%%%%%%%%
\begin{equation*}
A_{T,1}=\|\bar{\mathbf w}(i)\|^{2}\left[\|\mathbf r(i)\|^{2}-\frac{|\mathbf
r^{H}(i)\hat{\mathbf p}(i)|^{2}}{\|\hat{\mathbf p}(i)\|^{2}}\right].
\end{equation*}
By examining the second derivative of \eqref{eq:costfuncjirrlms11} with respect
to $\mu_{T}$, we conclude that $\mu_{T,1}$ and $\mu_{T,2}$ are the solutions
that correspond to the minima. In this work, the $\mu_{T,1}$ is used and a
positive real scaling factor $\mu_{T,0}$ is implemented that will not change
the direction of the tap-weight vector. Finally, the NSG update function of
$\mathbf{T}(i)$ is given by
%%%%%%%%%%%%%%%%%%%%%%%%%%%%%%%%%%%%%%%%%%%%%%%%%%%%%%%%%%%%%%%%%%%%%
\begin{equation}
\mathbf{T}(i+1)=\mathbf{T}(i)-y^{*}(i)\mu_{T,0}\mathbf
A_{T,2}-A_{T,3}\hat{\mathbf p}(i)\bar{\mathbf w}^{H}(i).\label{eq:rrnlmsTfinal}
\end{equation}where
%%%%%%%%%%%%%%%%%%%%%%%%%%%%%%%%%%%%%%%%%%%%%%%%%%%%%%%%%%%%%%%%%%%%%
\begin{equation*}
\mathbf A_{T,2}=\frac{|y(i)|-1}{|y(i)|A_{T,1}}\left(\mathbf r(i)\bar{\mathbf
w}^{H}(i)-\frac{\hat{\mathbf p}^{H}(i)\mathbf r(i)}{\|\hat{\mathbf
p}(i)\|^{2}}\hat{\mathbf p}(i)\bar{\mathbf w}^{H}(i)\right),
\end{equation*}
%%%%%%%%%%%%%%%%%%%%%%%%%%%%%%%%%%%%%%%%%%%%%%%%%%%%%%%%%%%%%%%%%%%%%
\begin{equation*}
A_{T,3}=\big(\|\bar{\mathbf w}(i)\|^{2}\|\hat{\mathbf
p}(i)\|^{2}\big)^{-1}\big(\hat{\mathbf p}^{H}(i)\mathbf{T}(i)\bar{\mathbf
w}(i)-v\big).
\end{equation*}
%%%%%%%%%%%%%%%%%%%%%%%%%%%%%%%%%%%%%%%%%%%%%%%%%%%%%%%%%%%%%%%%%%%%%
Now, let us adapt $\bar{\mathbf w}(i)$ while assuming $\mathbf{T}(i)$ is fixed.
The gradient of the Lagrangian with respect to $\bar{\mathbf w}(i)$ is given by
$\nabla_{w}\mathcal {L}_{\rm
JIO-NSG}=E\left[e(i)y^{*}(i)\mathbf{T}^{H}(i)\mathbf
r(i)\right]+\frac{1}{2}\lambda_{Nw}(i)\hat{\mathbf p}(i)\bar{\mathbf
w}^{H}(i)$, where $\lambda_{Nw}(i)$ is the complex-valued Lagrange multiplier
for updating the reduced-rank filter. By using the instantaneous estimator of
the gradient vector, the SG adaptation equation is given by
%%%%%%%%%%%%%%%%%%%%%%%%%%%%%%%%%%%%%%%%%%%%%%%%%%%%%%%%%%%%%%%%%%%%%
\begin{equation}
\bar{\mathbf w}(i+1)=\bar{\mathbf
w}(i)-\mu_{w}e(i)y^{*}(i)\mathbf{T}^{H}(i)\mathbf
r(i)-\mu_{w}\frac{\lambda_{Nw}(i)}{2}\mathbf{T}^{H}(i) \hat{\mathbf
p}(i).\label{eq:rrnlmswi1}
\end{equation}
%%%%%%%%%%%%%%%%%%%%%%%%%%%%%%%%%%%%%%%%%%%%%%%%%%%%%%%%%%%%%%%%%%%%%
Using the constraint $\bar{\mathbf w}^{H}(i+1) \mathbf T^{H}(i) \hat{\mathbf
p}(i)=v$, we have
%%%%%%%%%%%%%%%%%%%%%%%%%%%%%%%%%%%%%%%%%%%%%%%%%%%%%%%%%%%%%%%%%%%%%
\begin{equation}
\lambda_{Nw}(i)=2\frac{\hat{\mathbf p}^{H}(i)\mathbf T(i)\bar{\mathbf
w}(i)-\mu_{w} e(i)y^{*}(i)\hat{\mathbf
p}^{H}(i)\mathbf{T}(i)\mathbf{T}^{H}(i)\mathbf
r(i)-v}{\mu_{w}\|\mathbf{T}^{H}(i)\hat{\mathbf p}(i)\|^{2}}.
\label{eq:rrnlmslambdaw}
\end{equation}
%%%%%%%%%%%%%%%%%%%%%%%%%%%%%%%%%%%%%%%%%%%%%%%%%%%%%%%%%%%%%%%%%%%%%
The NSG algorithm for updating the reduced-rank filter aims at minimizing the
cost function
%%%%%%%%%%%%%%%%%%%%%%%%%%%%%%%%%%%%%%%%%%%%%%%%%%%%%%%%%%%%%%%%%%%%%
\begin{equation}
\mathbf J_{\rm JIO-NSG} (\mu_{w})=\frac{1}{2}\left[\left|\bar{\mathbf w}
^{H}(i+1) \mathbf T^{H}(i) \mathbf
r(i)\right|^{2}-1\right]^{2}.\label{eq:costfuncjirrlms2}
\end{equation}
Substituting \eqref{eq:rrnlmswi1} and \eqref{eq:rrnlmslambdaw} into
\eqref{eq:costfuncjirrlms2}, the solutions of $\mu_{w}$ that correspond to a
null gradient vector of \eqref{eq:costfuncjirrlms2} are given by
%%%%%%%%%%%%%%%%%%%%%%%%%%%%%%%%%%%%%%%%%%%%%%%%%%%%%%%%%%%%%%%%%%%%%
\begin{equation*}
\mu_{w,1}=\frac{|y(i)|-1}{|y(i)|e(i)A_{w,1}},~ \mu_{w,2}=\frac{|y(i)|+
1}{|y(i)|e(i)A_{w,1}},
\end{equation*}
%%%%%%%%%%%%%%%%%%%%%%%%%%%%%%%%%%%%%%%%%%%%%%%%%%%%%%%%%%%%%%%%%%%%%
\begin{equation*}
\mu_{w,3}=\mu_{w,4}=\frac{1}{e(i)A_{w,1}},
\end{equation*}where the scale term is given by
%%%%%%%%%%%%%%%%%%%%%%%%%%%%%%%%%%%%%%%%%%%%%%%%%%%%%%%%%%%%%%%%%%%%%%
%\begin{equation*}
%\begin{split}
%A_{w,1}&=\mathbf r^{H}(i)\mathbf{T}(i)\mathbf{T}^{H}(i)\mathbf
%r(i)\\
%&-\frac{\mathbf r^{H}(i)\mathbf{T}(i)\mathbf{T}^{H}(i)\hat{\mathbf
%p}(i)\hat{\mathbf p}^{H}(i)\mathbf{T}(i)\mathbf{T}^{H}(i)\mathbf
%r(i)}{\hat{\mathbf
%p}^{H}(i)\mathbf{T}(i)\mathbf{T}^{H}(i)\hat{\mathbf p}(i)}\\
%&=\parallel \mathbf{T}^{H}(i)\mathbf
%r(i)\parallel^{2}-\frac{|\mathbf
%r^{H}(i)\mathbf{T}(i)\mathbf{T}^{H}(i)\hat{\mathbf
%p}(i)|^{2}}{\parallel\mathbf{T}^{H}(i)\hat{\mathbf
%p}(i)\parallel^{2}}
%\end{split}
%\end{equation*}
%%%%%%%%%%%%%%%%%%%%%%%%%%%%%%%%%%%%%%%%%%%%%%%%%%%%%%%%%%%%%%%%%%%%%
\begin{equation*}
A_{w,1}=\| \mathbf{T}^{H}(i)\mathbf r(i)\|^{2}-\frac{|\mathbf
r^{H}(i)\mathbf{T}(i)\mathbf{T}^{H}(i)\hat{\mathbf
p}(i)|^{2}}{\|\mathbf{T}^{H}(i)\hat{\mathbf p}(i)\|^{2}}
\end{equation*}
By examining the second derivative of \eqref{eq:costfuncjirrlms2} with respect
to $\mu_{w}$, only $\mu_{w,1}$ and $\mu_{w,2}$ correspond to the minima of the
cost function \eqref{eq:costfuncjirrlms2}. Finally, by applying a positive real
scaling factor $\mu_{w,0}$ to control the tap-weight vector, the adaptation
equation by using $\mu_{w,1}$ is given by
%%%%%%%%%%%%%%%%%%%%%%%%%%%%%%%%%%%%%%%%%%%%%%%%%%%%%%%%%%%%%%%%%%%%%
\begin{equation}
\bar{\mathbf w}^{H}(i+1)=\bar{\mathbf
w}^{H}(i)-y^{*}(i)\mu_{w,0}\mathbf
A_{w,2}-A_{w,3}\mathbf{T}^{H}(i)\hat{\mathbf
p}(i).\label{eq:rrnlmswfinal}
\end{equation}where
%%%%%%%%%%%%%%%%%%%%%%%%%%%%%%%%%%%%%%%%%%%%%%%%%%%%%%%%%%%%%%%%%%%%%
\begin{equation*}
\mathbf A_{w,2}=\frac{|y(i)|-1}{|y(i)|A_{w,1}}\left(\mathbf{T}^{H}(i)\mathbf
r(i)-\frac{\hat{\mathbf p}^{H}(i)\mathbf{T}(i)\mathbf{T}^{H}(i)\mathbf
r(i)}{\|\mathbf{T}^{H}(i)\hat{\mathbf p}(i)\|^{2}}\mathbf{T}^{H}(i)\hat{\mathbf
p}(i)\right),
\end{equation*}
%%%%%%%%%%%%%%%%%%%%%%%%%%%%%%%%%%%%%%%%%%%%%%%%%%%%%%%%%%%%%%%%%%%%%
\begin{equation*}
A_{w,3}=\big(\|\mathbf{T}^{H}(i)\hat{\mathbf
p}(i)\|^{2}\big)^{-1}\big(\hat{\mathbf p}^{H}(i)\mathbf{T}(i)\bar{\mathbf
w}(i)-v\big).
\end{equation*}

In the proposed JIO-NSG scheme, $\mathbf T(i)$ and $\bar{\mathbf w}(i)$ are
computed jointly and iteratively. Let $c$ denotes the iteration number and
define $c_{max}$ as the total number of iterations for each time instant. We
have $\mathbf T_{0}(i)=\mathbf T_{c_{max}}(i-1)$ and $\bar{\mathbf
w}_{0}(i)=\bar{\mathbf w}_{c_{max}}(i-1)$. For the $c$-th iteration, $\mathbf
T_{c}(i)$ is updated with $\mathbf T_{c-1}(i)$ and $\bar{\mathbf w}_{c-1}(i)$
using \eqref{eq:rrnlmsTfinal}, then $\bar{\mathbf w}_{c}(i)$ is trained with
$\mathbf T_{c}(i)$ and $\bar{\mathbf w}_{c-1}(i)$ via \eqref{eq:rrnlmswfinal}.

It is interesting to note that the complexity of the JIO-NSG scheme
could be lower than the full-rank NSG algorithm because there are
many entries that are frequently reused in the update equations, for
example, the scalar term $\hat {\mathbf p}^{H}(i)\mathbf r(i)$, the
vectors of $\mathbf T^{H}(i)\hat {\mathbf p}(i)$ and $\mathbf
T^{H}(i)\mathbf r(i)$. However, the price we pay for the complexity
reduction is the requirement of extra storage space at the receiver.

\subsection{Blind Channel Estimator For the NSG Version}
\label{sec:bceforjionsg}
%The complexity of the blind channel estimator described in Section.
%\ref{sec:receiverdesign} is mainly loaded at \eqref{eq:hatV}.
For the JIO-NSG receiver, we rearrange the equation \eqref{eq:hatV}
as
%%%%%%%%%%%%%%%%%%%%%%%%%%%%%%%%%%%%%%%%%%%%%%%%%%%%%%%%%%%%%%%%%%%%%
\begin{equation}
\mathbf {\hat V}(i)=\mathbf {S}_{e}^{H}\mathbf P_{r}^{H}\mathbf
{\hat W}(i)\label{eq:hatVNSG}
\end{equation} where $\mathbf
{\hat W}(i)=\mathbf R^{-m}(i)\mathbf P_{r}\mathbf {S}_{e}$. Here, we
implement the Leakage SG algorithm to estimate $\mathbf {\hat
W}(i)$, which can be expressed as \cite{xgdoukopoulos2005}
%%%%%%%%%%%%%%%%%%%%%%%%%%%%%%%%%%%%%%%%%%%%%%%%%%%%%%%%%%%%%%%%%%%%%
\begin{equation}
\mathbf {\hat W}_{l}(i)=\lambda_{v}\mathbf {\hat
W}_{l}(i-1)+\mu_{v}(\mathbf {\hat W}_{l-1}(i)-\mathbf r(i)\mathbf
r^{H}(i)\mathbf {\hat W}_{l}(i-1)),\label{eq:hatW}
\end{equation} where $l=1,\dots,m$ is defined as the iteration
index, $\lambda_{v}$ is the leakage factor and $\mu_{v}$ is the step
size. Using \eqref{eq:hatVNSG}, we obtain the leakage SG blind
channel estimator that is given by
%%%%%%%%%%%%%%%%%%%%%%%%%%%%%%%%%%%%%%%%%%%%%%%%%%%%%%%%%%%%%%%%%%%%%
\begin{equation}
\mathbf {\hat h}(i)=\mathbf {\hat h}(i-1)-\left(\mathbf {\hat
V}(i)\mathbf {\hat h}(i-1)\right)/tr[\mathbf {\hat
V}(i)],\label{eq:hatHjionsg}
\end{equation} Finally, the effective signature vector
of the desired user is given by
\begin{equation}
\mathbf {\hat p}(i)=\mathbf P_{r}\mathbf {S}_{e}\mathbf {\hat h}(i)
\label{equ:pforjionsg}
\end{equation}
In terms of the computational complexity, we need $4mML$
multiplications and $3mML-mL$ additions for all the recursions in
\eqref{eq:hatW}; $L^{2}M$ multiplications and $L^{2}M-L^{2}$
additions for \eqref{eq:hatVNSG}.

The JIO-NSG version is summarized in Table. \ref{tab:adaptive}.
%%%%%%%%%%%%%%%%%%%%%%%%%%%%%%%%%%%%%%%%%%%%%%%%%%%%%%%%%%%%%%%%%%%%%
\linespread{1}
\begin{table}
\centering
 \caption{\normalsize Adaptive versions of the Proposed JIO Receiver.} {
\begin{tabular}{l}
\hline\hline\rule{0pt}{2.6ex} \rule[-1.2ex]{0pt}{0pt}
\bfseries {NSG version:}\\
\hline\rule{0pt}{2.6ex} \rule[-1.2ex]{0pt}{0pt}
\bfseries {Initialization:}  \\
$\bar {\mathbf w}(1)=[1,1,1,\dots,1]$, $D$-by-$1$ vector, \\
$\mathbf T (1)=[\mathbf I_{D} ~|~ \mathbf 0]^{T}$, $M$-by-$D$ matrix.\\
($\mathbf {I}_{D}$ represents the $D$-by-$D$ identity matrix.)\\

\hfill \\
{\it for} $i=1,2,\dots$\\
\hfill \\

\bfseries {1: Pre-adaptation:}\\
$\bar {\mathbf r}(i)=\mathbf {T}^{H}(i) \mathbf {r}(i)$, $y(i)=\mathbf {\bar w}^{H}(i)\bar {\mathbf r}(i)$,\\
Calculate $\mathbf {\hat V}(i)$ and $\mathbf {\hat h}(i)$ using
\eqref{eq:hatVNSG} and \eqref{eq:hatHjionsg}, respectively,\\
Calculate $\mathbf {\hat p}(i)$ using \eqref{equ:pforjionsg},\\
Set $\mathbf T_{0}(i+1)=\mathbf T_{c_{max}}(i)$ and $\bar {\mathbf
w}_{0}(i+1)=\bar {\mathbf w}_{c_{max}}(i)$.\\

\hfill \\
\bfseries {2: Adaptation of $\mathbf T(i+1)$ and $\bar {\mathbf w}(i+1)$:}\\
{\it for} $c=1,2,\dots,c_{max}$\\
Update $\mathbf T_{c}(i+1)$ using \eqref{eq:rrnlmsTfinal} with
$\mathbf T_{c-1}(i+1)$ and $\bar {\mathbf w}_{c-1}(i+1)$,\\
Update $\bar {\mathbf w}_{c}(i+1)$ using \eqref{eq:rrnlmswfinal}
with
$\mathbf T_{c}(i+1)$ and $\bar {\mathbf w}_{c-1}(i+1)$,\\
{\it end}\\
Set $\mathbf T(i+1)=\mathbf T_{c_{max}}(i+1)$ and $\bar {\mathbf
w}(i+1)=\bar {\mathbf w}_{c_{max}}(i+1)$\\
\hfill \\

\bfseries {4: Make Decision for the $i$-th data bit:}\\
%$y(i)=\bar {\mathbf{w}}^{H}(i)\bar {\mathbf{r}}(i)$\\
$\hat {b}(i)={\rm sign}(\mathfrak{R}(y(i)))$\\
 \hline\rule{0pt}{2.6ex} \rule[-1.2ex]{0pt}{0pt}
\bfseries {RLS version:}\\
\hline\rule{0pt}{2.6ex} \rule[-1.2ex]{0pt}{0pt}
\bfseries {Initialization:}  \\
$\bar {\mathbf w}(1)=[1,1,1,\dots,1]$, $D$-by-$1$ vector, \\
$\mathbf t_{d}(1)=[1,0,0,\dots,0]$ ($d=1,2,\dots,D$), $D$-by-$1$ vectors,\\
$\mathbf {\bar d}(0)=[0,0,\dots,0]$, $D$-by-$1$ vector,\\
$\mathbf {\hat R}_{y}^{-1}(0)=\mathbf {I}_{M}/\delta$, $M$-by-$M$ matrix,\\
$\mathbf {\hat R}_{\rm T}^{-1}(0)=\mathbf {I}_{D}/\delta$, $D$-by-$D$ matrix,\\
($\mathbf {I}_{M}$ is the $M$-by-$M$ identity matrix. $\delta$ is a positive constant.)\\
\hfill \\
{\it for} $i=1,2,\dots$\\
\hfill \\

\bfseries {1: Pre-adaptation:}\\
$\bar {\mathbf r}(i)=\mathbf {T}^{H}(i) \mathbf {r}(i)$, $y(i)=\mathbf {\bar w}^{H}(i)\bar {\mathbf r}(i)$,\\
$\mathbf {\bar d}(i)=\mathbf {\bar d}(i-1)+\alpha\mathbf {\bar
r}(i)y^{*}(i)$,\\
Estimate $\mathbf {\hat R}_{y}^{-1}(i)$ and $\mathbf {\hat R}_{\rm T}^{-1}(i)$ using \eqref{eq:CArlsinvR} and \eqref{eq:CArlsinvRt}, respectively,\\
Calculate $\mathbf {\hat V}(i)$ and $\mathbf {\hat h}(i)$ using
\eqref{eq:hatVjiorls} and \eqref{eq:hatHjiorls}, respectively,\\
Calculate $\mathbf {\hat p}(i)$ using \eqref{equ:pforjiorls}.\\

\hfill \\
\bfseries {2: Adaptation of $\mathbf t_{d}(i)$:}\\
{\it for} $d=1,2,\dotsb,D$\\
Calculate $\lambda_{t,d}(i)$ using \eqref{eq:CArlslambdt},\\
Update $\mathbf t_{d}(i)$ using \eqref{eq:CARLSupdatetd},\\
Normalize $\mathbf t_{d}(i)\leftarrow\mathbf
t_{d}(i)/\parallel\mathbf t_{d}(i)\parallel$.\\
\hfill \\

\bfseries {3: Adaptation of $\bar {\mathbf w}(i)$:}\\
Calculate $\lambda_{Rw}(i)$ using \eqref{eq:CArlslambdw},\\
Update $\bar {\mathbf w}(i)$ using \eqref{eq:carlsupdatew},\\
\hfill \\

\bfseries {4: Make Decision for the $i$-th data bit:}\\
%$y(i)=\bar {\mathbf{w}}^{H}(i)\bar {\mathbf{r}}(i)$\\
$\hat {b}(i)={\rm sign}(\mathfrak{R}(y(i)))$\\
 \hline
\end{tabular}}
\label{tab:adaptive}
\end{table}
\linespread{1.5}
%%%%%%%%%%%%%%%%%%%%%%%%%%%%%%%%%%%%%%%%%%%%%%%%%%%%%%%%%%%%%%%%%%%%%%%%%%%%%%%%%%%%%%%%%%%%%%%%%%%%%%%%%%%%%%%%%%%

%%%%%%%%%%%%%%%%%%%%%%%%%%%%%%%%%%%%%%%%%%%%%%%%%%%%%%%%%%%%%%%%%%%%%%%%%%%%%%%%%%%%%%%%%%%%%%%%%%%%%%%%%%%%%%%%%%%
%%%%%%%%%%%%%%%%%%%%%%%%%%%%%%%%%%%%%%%%%%%%%%%%%%%%%%%%%%%%%%%%%%%%%%%%%%%%%%%%%%%%%%%%%%%%%%%%%%%%%%%%%%%%%%%%%%%
\section{Proposed JIO-RLS Algorithms}
\label{sec:proposedjioca}

In this section we detail the RLS version of the proposed JIO scheme. In the
JIO scheme, the $M$-by-$D$ (where $D\ll M$) transformation matrix can be
expressed as
\begin{equation} \mathbf
T(i)=[\mathbf{t}_{1}(i),\mathbf{t}_{2}(i),\dots,\mathbf{t}_{D}(i)].
\end{equation}
Note that the reduced-rank received signal can be expressed as $\mathbf {\bar
r}(i)=\mathbf T^{H}(i)\mathbf r(i)$, whose $d$-th element is
$\bar{r}_{d}(i)=\mathbf{t}_{d}^{H}(i)\mathbf r(i)$. Since the transformation
matrix projects the received signal onto a small-dimensional subspace, these
vectors $\mathbf{t}_{d}(i)$ can be considered as the direction vectors on each
dimension of the subspace. For each time instant, we compute these
$M$-dimensional vectors $\mathbf{t}_{d}(i)$ (where $d=1,2,\dots,D$) one by one.
One of the advantages of this process method in the RLS version is that the
complexity of training the transformation matrix could be reduced with an
approximation which will be shown soon. In addition, this method provides a
better representation of the transformation matrix and leads to better
performance than the approach that updates all the columns of the projection
matrix together. It should be noted that, the NSG version can also be modified
to update the columns of the transformation matrix one by one, but the limited
improved performance in NSG version is not worth the payment of the increased
complexity.

After the projection, $\mathbf {\bar r}(i)$ is fed into the reduced-rank filter
$\bar{\mathbf{w}}(i)$ and the output signal is given by
%%%%%%%%%%%%%%%%%%%%%%%%%%%%%%%%%%%%%%%%%%%%%%%%%%%%%%%%%%%%%%%%%%%%%
\begin{equation*}
y(i)=\bar{\mathbf{w}}^{H}(i)\mathbf T^{H}(i)\mathbf r(i)=\bar{\mathbf w}
^{H}(i)\sum^{D}_{d=1}\mathbf{t}_{d}^{H}(i)\mathbf r(i)\mathbf q_{d},
\end{equation*} where $\mathbf q_{d}$ (where
$d=1,2,\dots,D$) are the vectors whose $d$-th elements are ones, while all the
other elements are zeros. In this section, an adaptive blind channel estimation
is employed and $\mathbf{t}_{d}(i)$ are optimized jointly and iteratively with
$\bar{\mathbf{w}}(i)$ via RLS algorithms.

%%%%%%%%%%%%%%%%%%%%%%%%%%%%%%%%%%%%%%%%%%%%%%%%%%%%%%%%%%%%%%%%%%%%%%%%%%%%%%%%%%%%%%%%%%%%%%%%%%%%%%%%%%%%%%%%%%%
%%%%%%%%%%%%%%%%%%%%%%%%%%%%%%%%%%%%%%%%%%%%%%%%%%%%%%%%%%%%%%%%%%%%%%%%%%%%%%%%%%%%%%%%%%%%%%%%%%%%%%%%%%%%%%%%%%%
\subsection{JIO-RLS Algorithms}
In the JIO-RLS scheme, we need to solve the optimization problem
%%%%%%%%%%%%%%%%%%%%%%%%%%%%%%%%%%%%%%%%%%%%%%%%%%%%%%%%%%%%%%%%%%%%%%%%%%%%%%%%%%%%%%%%%%%%%%%%%%%%%%%%
\begin{equation}
\left[{\mathbf{\bar
w}}(i),\mathbf{t}_{1}(i),\dots,\mathbf{t}_{D}(i)\right]=\mathop{\mbox{arg}}
\mathop{\mbox{min}}_{{\mathbf{w}}(i),\mathbf{t}_{1}(i),\dots,\mathbf{t}_{D}(i)}
\mathbf J_{JIO-RLS}
\big(\bar{\mathbf{w}}(i),\mathbf{t}_{1}(i),\dots,\mathbf{t}_{D}(i)\big),
\end{equation}subject to the constraint $\bar{\mathbf
w} ^{H}(i) \sum^{D}_{d=1}\mathbf{t}_{d}^{H}(i)\hat{\mathbf
p}(i)\mathbf q_{d}=v$, where $\hat{\mathbf p}(i)$ is the
estimated signature vector obtained via blind channel estimation
that will be detailed in Section \ref{sec:bceforjiorls}. $v$ is
a real-valued constant to ensure the convexity of the CM cost
function:
%%%%%%%%%%%%%%%%%%%%%%%%%%%%%%%%%%%%%%%%%%%%%%%%%%%%%%%%%%%%%%%%%%%%%
\begin{equation*}
\mathbf J_{JIO-RLS}
\big(\bar{\mathbf{w}}(i),\mathbf{t}_{1}(i),\dots,\mathbf{t}_{D}(i)\big)=\frac{1}{2}\sum^{i}_{j=1}\alpha^{i-j}\left(|y(j)|^{2}-1\right)^{2},
%\label{eq:CMcostfuncCARls}
\end{equation*} where $0<\alpha \leq 1$ is the forgetting
factor and $y(i)$ is the output signal at the $i$-th time instant. Let us now
consider the problem through the Lagrangian
\begin{equation}
\begin{split}
&\mathcal {L}_{JIO-RLS}\big(\bar{\mathbf{w}}(i),\mathbf{t}_{1}(i),\dots,\mathbf{t}_{D}(i)\big)=\frac{1}{2}\sum^{i}_{j=1}\alpha^{i-j}\left(\left|y(j)\right|^{2}-1\right)^{2}\\
&+\mathfrak{R}\left[\lambda_{R}(i)\big(\bar{\mathbf w}
^{H}(i)\sum^{D}_{d=1}\mathbf{t}_{d}^{H}(i)\mathbf {\hat p}(i)\mathbf
q_{d}-v\big)\right], \label{eq:CMcostfuncCARlsuncons}
\end{split}
\end{equation} where $\lambda_{R}(i)$ is a complex-valued Lagrange multiplier.
In the proposed JIO-RLS scheme, for each time instant, we firstly update the
vectors $\mathbf t_{d}(i)$ (where $d=1,2,\dots,D$) while assuming that
$\bar{\mathbf{w}}(i)$ and other column vectors are fixed. Then we adapt the
reduced-rank filter with the updated transformation matrix.

%Now, let us define the output signal as $y(i)=\bar{\mathbf w}
%^{H}(i)\mathbf{\bar r}(i)$, where the reduced-rank signal can be
%expressed as $\mathbf{\bar
%r}(i)=\sum^{D}_{d=1}\mathbf{t}_{d}^{H}(i)\mathbf r(i)\mathbf q_{d}$.
For the update of the column vectors of the transformation matrix, we can
express the output signal as follows
%%%%%%%%%%%%%%%%%%%%%%%%%%%%%%%%%%%%%%%%%%%%%%%%%%%%%%%%%%%%%%%%%%%%%
\begin{equation*}
y(i)=\bar{\mathbf w} ^{H}(i)\sum^{D}_{d=1}\mathbf{t}_{d}^{H}(i)\mathbf
r(i)\mathbf q_{d}=\bar{w}_{d}^{*}(i)\bar{r}_{d}(i)+\bar{\mathbf w}
^{H}(i)\mathbf {\bar r}_{e}(i),
\end{equation*} where the $D$-dimensional vector $\mathbf {\bar r}_{e}(i)$ can be obtained by
calculating the reduced-rank received signal $\mathbf{\bar r}(i)$ and setting
its $d$-th element to zero. By taking the gradient term of
\eqref{eq:CMcostfuncCARlsuncons} with respect to $\mathbf t_{d}(i)$ and setting
it to a null vector, we have $\nabla_{t_{d}}\mathcal
{L}_{JIO-RLS}$$=\sum^{i}_{j=1}\alpha^{i-j}e(j)\mathbf r(j)$$\big(|\bar
{w}_{d}(j)|^{2}\mathbf r^{H}(j)\mathbf t_{d}(i)$$+\bar {w}_{d}^{*}(j)\mathbf
{\bar r}_{e}^{H}(j)\bar{\mathbf w}(j)\big)$ $+\frac{1}{2}\lambda_{t,d}(i) \bar
{w}_{d}^{*}(i)\mathbf {\hat p}(i)=0$, where $e(i)=|y(i)|^{2}-1$ and
$\lambda_{t,d}(i)$ is the complex-valued Lagrange multiplier for updating the
$d$-th column vector in the transformation matrix. Rearranging the terms we
obtain
%%%%%%%%%%%%%%%%%%%%%%%%%%%%%%%%%%%%%%%%%%%%%%%%%%%%%%%%%%%%%%%%%%%%%
\begin{equation}
\mathbf t_{d}(i)=-\mathbf R_{\rm d}^{-1}(i)\left(\frac{\lambda_{t,d}(i)}{2}
\bar {w}_{d}^{*}(i)\mathbf {\hat p}(i)+\mathbf v_{r}(i)\right),
\end{equation} where we define the $M$-dimensional vector $\mathbf v_{r}(i)=\sum^{i}_{j=1}\alpha^{i-j}\bar
{w}_{d}^{*}(j)\mathbf r(j)\big(e(j)\mathbf r_{e}^{H}(j)\bar{\mathbf
w}(j)-\bar {w}_{d}(j)\bar{r}_{d}^{*}(j)\big)$ and the $M$-by-$M$
matrix $\mathbf R_{\rm d}(i)=\sum^{i}_{j=1}\alpha^{i-j}|\bar
{w}_{d}(j)|^{2}|y(j)|^{2}\mathbf r(j)\mathbf r^{H}(j)$. Note that,
$\mathbf R_{\rm d}(i)$ is dependent on $\bar {w}_{d}(i)$, which is
the $d$-th element of the reduced-rank filter. Hence, for updating
each $\mathbf t_{d}(i)$, we need to calculate the corresponding
$\mathbf R_{\rm d}^{-1}(i)$ and that leads to high computational
complexity. In our work, we devise an approximation $\mathbf R_{\rm
d}(i)\approx|\bar
{w}_{d}(i)|^{2}\sum^{i}_{j=1}\alpha^{i-j}|y(j)|^{2}\mathbf
r(j)\mathbf r^{H}(j)=|\bar {w}_{d}(i)|^{2}\mathbf R_{y}(i)$. Then we
adopt the matrix inversion lemma \cite{haykin} to recursively
estimate $\mathbf R_{y}^{-1}(i)$ as follows
%%%%%%%%%%%%%%%%%%%%%%%%%%%%%%%%%%%%%%%%%%%%%%%%%%%%%%%%%%%%%%%%%%%%%
\begin{equation}
\begin{split}
&\bm {\kappa}_{y} (i)=\mathbf {\hat R}_{y}^{-1}(i-1)y(i)\mathbf r(i),\\
&\phi_{y}(i)=\frac{1}{\alpha + y^{*}(i)\mathbf r^{H}(i)\bm {\kappa}_{y} (i)},\\
&\mathbf {\hat R}_{y}^{-1}(i)=\frac{1}{\alpha}\left(\mathbf {\hat R}_{y}^{-1}(i-1)-(\phi(i)\bm {\kappa}_{y}(i))\bm {\kappa}_{y}^{H}(i) \right),\\
\end{split}\label{eq:CArlsinvR}
\end{equation} where $\mathbf {\hat R}_{y}^{-1}(i)$ is the estimate of $\mathbf
{R}_{y}^{-1}(i)$. We use $\mathbf {\hat R}_{y}^{-1}(i)$ for all the
adaptations of $\mathbf t_{d}(i)$ to avoid the estimation of the
$\mathbf R_{\rm d}^{-1}(i)$ (where $d=1,2,\dots,D$) and the new
update equation is given by
%%%%%%%%%%%%%%%%%%%%%%%%%%%%%%%%%%%%%%%%%%%%%%%%%%%%%%%%%%%%%%%%%%%%%
\begin{equation}
\mathbf t_{d}(i)=-\frac{\mathbf {\hat R}_{y}^{-1}(i)}{|\bar
{w}_{d}(i)|^{2}}\left(\frac{\lambda_{t,d}(i)}{2} \bar {w}_{d}^{*}(i)\mathbf
{\hat p}(i)+\mathbf v_{r}(i)\right). \label{eq:CARLSupdatetd}
\end{equation}
Using the constraint $\bar{\mathbf w} ^{H}(i)
\sum^{D}_{d=1}\mathbf{t}_{d}^{H}(i)\hat{\mathbf p}(i)\mathbf q_{d}=v$, we
obtain the expression of the Lagrange multiplier as
%%%%%%%%%%%%%%%%%%%%%%%%%%%%%%%%%%%%%%%%%%%%%%%%%%%%%%%%%%%%%%%%%%%%%
\begin{equation}
\lambda_{t,d}(i)= 2\left[\frac{\bar {w}_{d}^{*}(i)\mathbf v_{r}^{H}(i)\mathbf
{\hat R}_{y}^{-1}(i)\mathbf {\hat p}(i)+\big(v-\bar {\mathbf
w}^{H}(i)\mathbf {\hat p}_{d}(i)\big)|\bar {w}_{d}(i)|^{2}}{-|\bar
{w}_{d}(i)|^{2}\mathbf {\hat p}^{H}(i)\mathbf {\hat R}_{y}^{-1}(i)\mathbf {\hat
p}(i)}\right]^{*}, \label{eq:CArlslambdt}
\end{equation} where $\mathbf{\hat p}_{d}(i)$ can be obtained by
calculating the vector $\mathbf T^{H}(i)\mathbf {\hat p}(i)$ and setting its
$d$-th element to zero. Note that in the update equation
\eqref{eq:CARLSupdatetd}, small values of $|\bar {w}_{d}(i)|^{2}$ may cause
numerical problems for the later calculation. This issue can be addressed by
normalizing the column vector after each adaptation, which is given by $\mathbf
t_{d}(i)\leftarrow\mathbf t_{d}(i)/\|\mathbf t_{d}(i)\|$.

After updating the transformation matrix column by column, now we are going to
adapt the reduced-rank filter $\bar{\mathbf w}(i)$. By assuming that the
transformation matrix is fixed, we can express the output signal in a simpler
way as
%%%%%%%%%%%%%%%%%%%%%%%%%%%%%%%%%%%%%%%%%%%%%%%%%%%%%%%%%%%%%%%%%%%%%
\begin{equation}
y(i)=\bar{\mathbf w} ^{H}(i)\mathbf T^{H}(i)\mathbf r(i),
\label{eq:outputsignalCARlssimple}
\end{equation} where $\mathbf
T(i)=[\mathbf{t}_{1}(i),\dots,\mathbf{t}_{D}(i)]$ and the constraint can be
expressed as $\bar{\mathbf w} ^{H}(i) \mathbf T^{H}(i)\hat{\mathbf
p}(i)=v$. Hence, the Lagrangian becomes
%%%%%%%%%%%%%%%%%%%%%%%%%%%%%%%%%%%%%%%%%%%%%%%%%%%%%%%%%%%%%%%%%%%%%%%%%%%%%%%%%%%%%%%%%%%%%%%%%%%%%%%%
\begin{equation}
\mathcal {L}_{JIO-RLS}\big(\bar{\mathbf{w}}(i),\mathbf
T(i)\big)=\frac{1}{2}\sum^{i}_{j=1}\alpha^{i-j}\left(\left|y(j)\right|^{2}-1\right)^{2}+\mathfrak{R}[\lambda_{R}(i)\big(\bar{\mathbf
w} ^{H}(i)\mathbf T^{H}(i)\mathbf {\hat p}(i)-v\big)].
\label{eq:CMcostfuncCARlsunconssimple}
\end{equation}
By taking the gradient term of \eqref{eq:CMcostfuncCARlsunconssimple} with
respect to $\bar{\mathbf{w}}(i)$ and setting it to a null vector, we have
$\nabla_{{\mathbf{w}}}\mathcal {L}_{JIO-RLS}
=\sum^{i}_{j=1}\alpha^{i-j}e(j)\mathbf T^{H}(j)\mathbf r(j)\mathbf
r^{H}(j)\mathbf T(j)\bar{\mathbf{w}}(i)$+$\frac{1}{2}\lambda_{Rw}(i)\mathbf
T^{H}(i)\mathbf {\hat p}(i)$$=0$, where the real-valued error is
$e(i)=\left(|y(i)|^{2}-1 \right)$ and $\lambda_{Rw}(i)$ is the complex-valued
Lagrange multiplier for updating the reduced-rank filter, rearranging the terms
we obtain
%%%%%%%%%%%%%%%%%%%%%%%%%%%%%%%%%%%%%%%%%%%%%%%%%%%%%%%%%%%%%%%%%%%%%%%%%%%%%%%%%%%%%%%%%%%%%%%%%%%%%%%%
\begin{equation}
\bar{\mathbf{w}}(i)=\mathbf R_{\rm T}^{-1}(i)\left(
-\frac{\lambda_{Rw}(i)}{2}\mathbf T^{H}(i)\mathbf {\hat p}(i)+\mathbf {\bar
d}(i)\right),\label{eq:carlsupdatew}
\end{equation} where $\mathbf R_{\rm T}(i)=\sum^{i}_{j=1}\alpha^{i-j}|y(j)|^{2}\mathbf {\bar r}(j)\mathbf {\bar
r}^{H}(j)$ and $\mathbf {\bar d}(i)=\sum^{i}_{j=1}\alpha^{i-j}\mathbf {\bar
r}(j)y^{*}(j)=\mathbf {\bar d}(i-1)+\alpha\mathbf {\bar r}(i)y^{*}(i)$. The
matrix inversion lemma \cite{haykin} is used again to recursively estimate the
inversion matrix $\mathbf R^{-1}_{\rm T}(i)$ as follows
%%%%%%%%%%%%%%%%%%%%%%%%%%%%%%%%%%%%%%%%%%%%%%%%%%%%%%%%%%%%%%%%%%%%%
\begin{equation}
\begin{split}
&\bm {\kappa}_{\rm T} (i)=\mathbf {\hat R}_{\rm T}^{-1}(i-1)\mathbf {\bar r}(i)y(i),\\
&\phi_{\rm T}(i)=\frac{1}{\alpha + y(i)^{*}\mathbf {\bar r}^{H}(i)\bm {\kappa}_{\rm T} (i)},\\
&\mathbf {\hat R}^{-1}_{\rm T}(i)=\frac{1}{\alpha}\left(\mathbf {\hat R}^{-1}_{\rm T}(i-1)-(\phi_{\rm T}(i)\bm {\kappa}_{\rm T}(i))\bm {\kappa}_{\rm T}^{H}(i) \right),\\
\end{split}\label{eq:CArlsinvRt}
\end{equation} where $\mathbf {\hat R}^{-1}_{\rm T}(i)$ is the
estimate of $\mathbf {R}^{-1}_{\rm T}(i)$. For calculating the Lagrange
multiplier, we use the constraint $\bar{\mathbf w} ^{H}(i)\mathbf
T^{H}(i)\mathbf {\hat p}(i)=v$ and obtain
%%%%%%%%%%%%%%%%%%%%%%%%%%%%%%%%%%%%%%%%%%%%%%%%%%%%%%%%%%%%%%%%%%%%%
\begin{equation}
\lambda_{Rw}(i)=2\left[\frac{\mathbf {\bar d}^{H}(i)\mathbf R_{\rm
T}^{-1}(i)\mathbf T^{H}(i)\mathbf {\hat p}(i)-v}{\mathbf {\hat
p}^{H}(i)\mathbf T(i)\mathbf R_{\rm T}^{-1}(i)\mathbf T^{H}(i)\mathbf {\hat
p}(i)}\right]^{*}. \label{eq:CArlslambdw}
\end{equation}

%%%%%%%%%%%%%%%%%%%%%%%%%%%%%%%%%%%%%%%%%%%%%%%%%%%%%%%%%%%%%%%%%%%%%%%%%%%%%%%%%%%%%%%%%%%%%%%%%%%%%%%%%%%%%%%%%%%
%%%%%%%%%%%%%%%%%%%%%%%%%%%%%%%%%%%%%%%%%%%%%%%%%%%%%%%%%%%%%%%%%%%%%%%%%%%%%%%%%%%%%%%%%%%%%%%%%%%%%%%%%%%%%%%%%%%
\subsection{Blind Channel Estimator For the RLS version}
\label{sec:bceforjiorls}

In the JIO-RLS algorithm, the estimation of the covariance matrix $\mathbf
R_{y}(i)=\sum^{i}_{j=1}\alpha^{i-j}|y(j)|^{2}\mathbf r(j)\mathbf r^{H}(j)$ and
its inversion are obtained in the stage of adapting the transformation matrix.
It should be noted that $|y(j)|^{2}$ tends to 1 as the number of received
signal increasing. Hence, by replacing the inverse matrix $\mathbf R^{-1}(i)$
in \eqref{eq:hatV} with $\mathbf R_{y}^{-1}(i)$, we obtain
%%%%%%%%%%%%%%%%%%%%%%%%%%%%%%%%%%%%%%%%%%%%%%%%%%%%%%%%%%%%%%%%%%%%%
\begin{equation}
\mathbf {\hat h}(i)=\mathbf {\hat h}(i-1)-\left(\mathbf {\hat
V}(i)\mathbf {\hat h}(i-1)\right)/tr[\mathbf {\hat
V}(i)],\label{eq:hatHjiorls}
\end{equation} where the $L$-by-$L$ matrix is defined as
%%%%%%%%%%%%%%%%%%%%%%%%%%%%%%%%%%%%%%%%%%%%%%%%%%%%%%%%%%%%%%%%%%%%%
\begin{equation}
\mathbf {\hat V}(i)=\mathbf {S}_{e}^{H}\mathbf P_{r}^{H}\mathbf
R_{y}^{-m}(i)\mathbf P_{r}\mathbf {S}_{e},\label{eq:hatVjiorls}
\end{equation} and the effective signature vector of the desired
user is given by
\begin{equation}
\mathbf {\hat p}(i)=\mathbf P_{r}\mathbf {S}_{e}\mathbf {\hat h}(i)
\label{equ:pforjiorls}
\end{equation}
Using $\mathbf R_{y}^{-1}(i)$ instead of $\mathbf R^{-1}(i)$ can save $\mathcal
{O} (M^2)$ computational complexity for the JIO-RLS version and simulation
results will demonstrate later that the performance will not be degraded with
this replacement.
%%%%%%%%%%%%%%%%%%%%%%%%%%%%%%%%%%%%%%%%%%%%%%%%%%%%%%%%%%%%%%%%%%%%%
The JIO-RLS version is summarized in Table. \ref{tab:adaptive}.
%%%%%%%%%%%%%%%%%%%%%%%%%%%%%%%%%%%%%%%%%%%%%%%%%%%%%%%%%%%%%%%%%%%%%

%%%%%%%%%%%%%%%%%%%%%%%%%%%%%%%%%%%%%%%%%%%%%%%%%%%%%%%%%%%%%%%%%%%%%%%%%%%%%%%%%%%%%%%%%%%%%%%%%%%%%%%%%%%%%%%%%%%
%%%%%%%%%%%%%%%%%%%%%%%%%%%%%%%%%%%%%%%%%%%%%%%%%%%%%%%%%%%%%%%%%%%%%%%%%%%%%%%%%%%%%%%%%%%%%%%%%%%%%%%%%%%%%%%%%%%
\section{Complexity analysis and Rank adaptation algorithm}
\label{sec:complexityanalysis}

In this section, a complexity analysis is presented to compare the two versions
of the JIO receiver, the full-rank NSG and RLS schemes, the NSG and RLS
versions of the MSWF. The computational complexity of the blind channel
estimators that are implemented in this work are also analyzed. A rank
adaptation algorithm is detailed in this section which is able to select the
rank adaptively and can achieve better tradeoffs between the convergence speed
and the steady state performances.

\subsection{Complexity analysis}

%%%%%%%%%%%%%%%%%%%%%%%%%%%%%%%%%%%%%%%%%%%%%%%%%%%%%%%%%%%%%%%%%%%%%%%%%%%%%%%%%%%%%%%%%%%%%%%%%%%%%%%%
\begin{table*} \centering \caption{\normalsize Complexity
analysis}
\begin{tabular}{l l l}
\hline\hline   & Complex Additions & Complex Multiplications \rule{0pt}{2.6ex} \rule[-1.2ex]{0pt}{0pt}\\
 \hline
%Full-Rank SG      & $3M+2$                      & $4M+3$ \rule{0pt}{2.6ex}\\
Full-Rank NSG     & $M^{2}+3M-1$                & $2M^{2}+4M+5$\rule{0pt}{2.6ex}\\
Full-Rank RLS     & $5M^{2}+2M+1$               & $5M^{2}+3M+1$ \\
MSWF-NSG          & $DM^{2}+(2D+2)M-2D^{2}-2$   & $(D+1)M^{2}+(4D+2)M-2D^{2}+4D+5$ \\
MSWF-RLS          & $DM^{2}+(2D+2)M+2D^{2}-D$   & $(D+1)M^{2}+(4D+2)M+2D^{2}+3D+1$ \\
JIO-NSG           & $c_{max}(6DM+3M+4D-2)$      & $c_{max}(8DM+4M+7D+11)$\\
JIO-RLS           & $DM^{2}+3DM+4D^{2}-4D$      & $DM^{2}+6DM+4D^{2}+15D+1$\\
 \hline
Conventional BCE  & $(m+1)M^{2}L-(m+1)ML+2M^{2}+L^{2}+L-1$ &$(m+1)M^{2}L+3M^{2}+L^{2}+2M+1$\rule{0pt}{2.6ex}\\
BCE for JIO-NSG   & $L^{2}M+3mML-(m-1)L-1$          & $L^{2}M+4mML+L^{2} $ \\
BCE for JIO-RLS   & $(m+1)M^{2}L-(m+1)ML+L^{2}+L-1$ & $(m+1)M^{2}L+L^{2}$ \\
 \hline
\end{tabular}
\label{tab:Complexity analysis}
\end{table*}
%%%%%%%%%%%%%%%%%%%%%%%%%%%%%%%%%%%%%%%%%%%%%%%%%%%%%%%%%%%%%%%%%%%%%%%%%%%%%%%%%%%%%%%%%%%%%%%%%%%%%%%%
As shown in Table. \ref{tab:Complexity analysis}, the complexity of the
analyzed blind CCM full-rank NSG and RLS, MSWF-NSG and MSWF-RLS
\cite{Rodrigo2008add} and the proposed NSG and RLS versions of the JIO scheme
is compared with respect to the number of complex additions and complex
multiplications for each time instant. The complexity of the conventional blind
channel estimator (BCE) that is described in Section \ref{sec:receiverdesign}
is compared with the BCEs for the JIO-NSG and JIO-RLS that are described in
Section \ref{sec:bceforjionsg} and Section \ref{sec:bceforjiorls},
respectively.

\begin{figure}[htb]
\begin{minipage}[h]{1.0\linewidth}
  \centering
  \centerline{\epsfig{figure=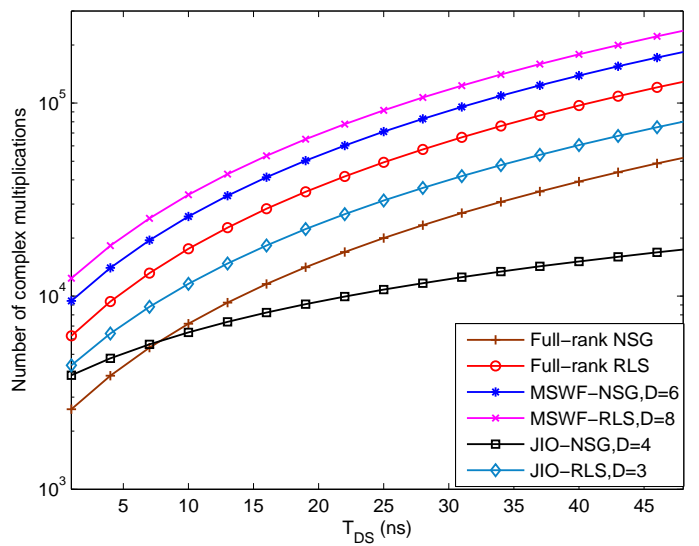,scale=0.7}}
\end{minipage}
\caption{Number of multiplications required for different algorithms.}
\label{fig:complexalgorithm}
\end{figure}

For the analysis of the adaptive algorithms, the quantity $M$ is the length of
the full-rank filter, $D$ is the dimension of reduced-rank filter and $c_{max}$
is the number of iterations for the JIO-NSG version in each time instant. Note
that, only one iteration is required in the JIO-RLS version for each time
instant. For the analysis of the BCEs, the quantity $L$ is the length of the
CIR and $m$ is the power of the inverse covariance matrix. In this work, $M$ is
the minimum integer that is larger than the scalar term
$(T_{s}/T_{\tau}+T_{DS}/T_{\tau}-1)/(T_{c}/T_{\tau})=(T_{s}+T_{DS}-T_{\tau})/T_{c}$
and $L=T_{DS}/T_{\tau}$. Since $T_{\tau}$ is set to $0.125ns$ as for the
standard IEEE802.15.4a channel model, symbol duration $T_{s}$ and chip duration
$T_{c}$ are assumed given for the designer. Hence, $M$ and $L$ are both related
to the channel delay spread $T_{DS}$. In this work, the parameters are set as
follows: $T_{s}=12ns$, $T_{c}=0.375ns$, $m=3$ and $c_{max}=3$. As shown in Fig.
\ref{fig:complexalgorithm}, the number of complex multiplications required for
different algorithms are compared as a function of the channel delay spread
$T_{DS}$. The JIO-RLS algorithm with $D=3$ has lower complexity than the MSWF
algorithms and the full-rank RLS. It will be demonstrated by the simulation
results that the JIO-RLS algorithm can achieve fast convergence with a very
small rank ($D<5$). The proposed JIO-NSG algorithm has lower complexity than
the full-rank NSG algorithm in the long channel delay spread scenarios. As
discussed in Section \ref{sec:jionsg}, the price we pay for such a complexity
reduction is the extra storage space at the receiver.

\begin{figure}[htb]
\begin{minipage}[h]{1.0\linewidth}
  \centering
  \centerline{\epsfig{figure=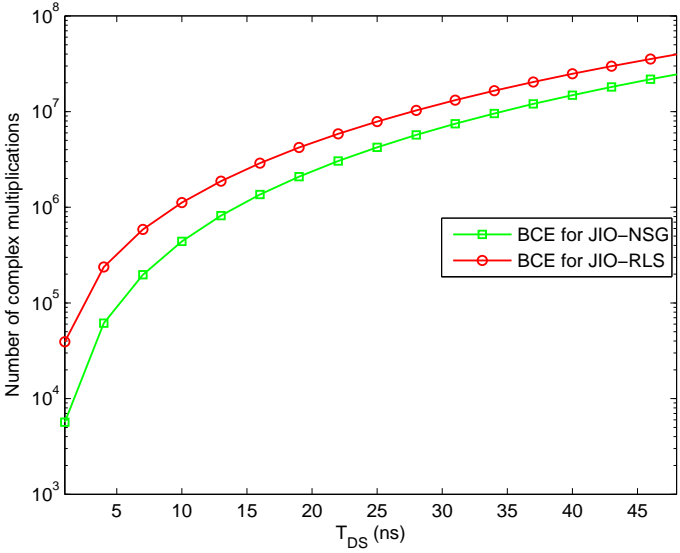,scale=0.7}}
\end{minipage}
\caption{Number of multiplications required for BCEs.} \label{fig:complexbce}
\end{figure}

The complexity of the BCEs for the JIO versions is shown in Fig.
\ref{fig:complexbce}, in which the number of complex multiplications is shown
as a function of channel delay spread $T_{DS}$. The complexity of the BCE for
the JIO-NSG version has lower complexity than the BCE for the JIO-RLS version
in all the analyzed scenarios.

%%%%%%%%%%%%%%%%%%%%%%%%%%%%%%%%%%%%%%%%%%%%%%%%%%%%%%%%%%%%%%%%%%%%%
%%%%%%%%%%%%%%%%%%%%%%%%%%%%%%%%%%%%%%%%%%%%%%%%%%%%%%%%%%%%%%%%%%%%%
%%%%%%%%%%%%%%%%%%%%%%%%%%%%%%%%%%%%%%%%%%%%%%%%%%%%%%%%%%%%%%%%%%%%%
%%%%%%%%%%%%%%%%%%%%%%%%%%%%%%%%%%%%%%%%%%%%%%%%%%%%%%%%%%%%%%%%%%%%%

\subsection{Rank Adaptation}
\label{sec:rankadaptation}

In the proposed blind JIO reduced-rank receiver, the computational complexity
and the performance are sensitive to the determined rank $D$. In this section,
a rank adaptation algorithm is employed to achieve better tradeoffs between the
performance and the complexity of the JIO receiver. The rank adaptation
algorithm is based on the \textit{a posteriori} LS cost function to estimate
the MSE, which is a function of $\bar {\mathbf{w}}_{D}(i)$ and $\mathbf
T_{D}(i)$ and can be expressed as
%%%%%%%%%%%%%%%%%%%%%%%%%%%%%%%%%%%%%%%%%%%%%%%%%%%%%%%%%%%%%%%%%%%%%%%%%%%%%%%%%%%%%%%%%%%%%%%%%%%%%%%%
\begin{equation}
\mathscr{C}_{D}(i)=\sum^{i}_{n=0} \lambda_{D}^{i-n}\left(|\bar {\mathbf
w}_{D}^{H}(n)\mathbf T_{D}^{H}(n)\mathbf r(n)|^{2}-1\right)^{2},
\label{eq:rankadaptationso}
\end{equation} where $\lambda_{D}$ is a forgetting factor.
Since the optimal rank can be considered as a function of the time interval $i$
\cite{MLHonig2002}, the forgetting factor is required and allows us to track
the optimal rank. For each time instant, we update a transformation matrix
$\mathbf T_{\rm M}(i)$ and a reduced-rank filter $\bar {\mathbf w}_{\rm M}(i)$
with the maximum rank $D_{\rm max}$, which can be expressed as
%%%%%%%%%%%%%%%%%%%%%%%%%%%%%%%%%%%%%%%%%%%%%%%%%%%%%%%%%%%%%%%%%%%%%%%%%%%%%%%%%%%%%%%%%%%%%%%%%%%%%%%%
\begin{equation}
\begin{split}
& \mathbf T_{\rm M}(i)=[\mathbf t_{\rm M,1}(i), \dots, \mathbf
t_{\rm M,\it D}(i), \dots, \mathbf t_{\rm M,\it
D_{\rm max}}(i)]^{T}\\
& \bar {\mathbf w}_{\rm M}(i)=[\bar {w}_{\rm M,1}(i), \dots, \bar
{w}_{\rm M,\it D}(i), \dots, \bar{w}_ {\rm M,\it D_{\rm
max}}(i)]^{T}\\
\end{split}
\end{equation}
After the adaptation, we test values of $D$ within the range $D_{\rm
min}$ to $D_{\rm max}$. For each tested rank, we use the following
estimators
%%%%%%%%%%%%%%%%%%%%%%%%%%%%%%%%%%%%%%%%%%%%%%%%%%%%%%%%%%%%%%%%%%%%%%%%%%%%%%%%%%%%%%%%%%%%%%%%%%%%%%%%
\begin{equation}
\begin{split}
& \mathbf T_{D}(i)=[\mathbf t_{\rm M,1}(i), \dots, \mathbf t_{\rm M,\it D}(i)]^{T}\\
& \bar {\mathbf w}_{D}(i)=[\bar {w}_{\rm M,1}(i), \dots,
\bar {w}_{\rm M,\it D}(i)]^{T}\\
\end{split}
\label{eq:dadaptparameters}
\end{equation}
%%%%%%%%%%%%%%%%%%%%%%%%%%%%%%%%%%%%%%%%%%%%%%%%%%%%%%%%%%%%%%%%%%%%%%%%%%%%%%%%%%%%%%%%%%%%%%%%%%%%%%%%
and substitute \eqref{eq:dadaptparameters} into
\eqref{eq:rankadaptationso} to obtain the value of
$\mathscr{C}_{D}(i)$, where $D\in\{D_{\rm min},\ldots,D_{\rm
max}\}$. The proposed algorithm can be expressed as
%%%%%%%%%%%%%%%%%%%%%%%%%%%%%%%%%%%%%%%%%%%%%%%%%%%%%%%%%%%%%%%%%%%%%%%%%%%%%%%%%%%%%%%%%%%%%%%%%%%%%%%%
\begin{equation}
D_{\rm opt}(i)=\arg \min_{D\in\{D_{\rm min},\ldots,D_{\rm
max}\}}\mathscr{C}_{D}(i). \label{eq:rankadp}
\end{equation}
%%%%%%%%%%%%%%%%%%%%%%%%%%%%%%%%%%%%%%%%%%%%%%%%%%%%%%%%%%%%%%%%%%%%%%%%%%%%%%%%%%%%%%%%%%%%%%%%%%%%%%%%
We remark that the complexity of updating the reduced-rank filter and the
transformation matrix in the proposed rank adaptation algorithm is the same as
the receiver with rank $D_{\rm max}$, since we only adapt the $\mathbf T_{\rm
M}(i)$ and $\bar {\mathbf w}_{\rm M}(i)$ for each time instant. However,
additional computations are required for calculating the values of
$\mathscr{C}_{D}(i)$ and selecting the minimum value of a $(D_{\rm max}-D_{\rm
min}+1)$-dimensional vector that corresponds to a simple search and comparison.

%%%%%%%%%%%%%%%%%%%%%%%%%%%%%%%%%%%%%%%%%%%%%%%%%%%%%%%%%%%%%%%%%%%%%%%%%%%%%%%%%%%%%%%%%%%%%%%%%%%%%%%%%%%%%%%%%%%
\section{Simulations}
\label{sec:simulations}

In this section, the proposed NSG and RLS versions of the blind JIO
adaptive receivers are applied to the uplink of a multiuser BPSK
DS-UWB system. The performance of the proposed receivers are
compared with the RAKE receiver with the maximal-ratio combining
(MRC), the NSG and RLS versions of the full-rank schemes and the
MSWF. Note that, the blind channel estimation described in section
\ref{sec:receiverdesign} is implemented to provide channel
coefficients to the RAKE receiver and its bit-error rate (BER)
performance is averaged for the purpose of comparison. In all
simulations, all the users are assumed to be transmitting
continuously at the same power level. The pulse shape adopted is the
RRC pulse with the pulse-width $0.375$ns. The spreading codes are
generated randomly for each user with a spreading gain of $32$ and
the data rate of the communication is approximately $83$Mbps. {  We
assess the blind receivers with the standard IEEE 802.15.4a channel
models of channel model 1 (ChMo1) and channel model 2 (ChMo2), which
are for indoor residential line-of-sight (LOS) and non-line of sight
(NLOS) environments, respectively. \cite{Molisch2005}.} We assume
that the channel is constant during the whole transmission. The
channel delay spread is $T_{DS}$ = 10ns and the ISI from 2 neighbor
symbols are taken into account for the simulations. The sampling
rate at the receiver is assumed to be 2.67GHz and the length of the
discrete time received signal is M = 59. For all the experiments,
all the adaptive receivers are initialized as vectors with all the
elements equal to 1. This allows a fair comparison between the
analyzed techniques for their convergence performance. In practice,
the filters can be initialized with prior knowledge about the
spreading code or the channel to accelerate the convergence. In all
the simulations, the phase of $h(0)$ is used as a reference to
remove the phase ambiguity derived from the blind channel estimates.
All the curves shown in this section are obtained by averaging 200
independent runs. { In this section, the coded bit error rate (BER)
performances are obtained by adopting a convolutional code with a
coding-rate of 2/3. The code polynomial is [7,5,5] and the
constraint length is set to 5. It should be noted that other coding
schemes employing Turbo codes, LDPC codes and/or iterative detection
\cite{Rodrigo2008DF} can be employed to further improve the
performance of the proposed algorithms.}

%%%%%%%%%%%%%%%%%%%%%%%%%%%%%%%%%%%%%%%%%%%%%%%%%%%%%%%%%%%%%%%%%%%%%
\begin{figure}[htb]
\begin{minipage}[h]{1.0\linewidth}
  \centering
  \centerline{\epsfig{figure=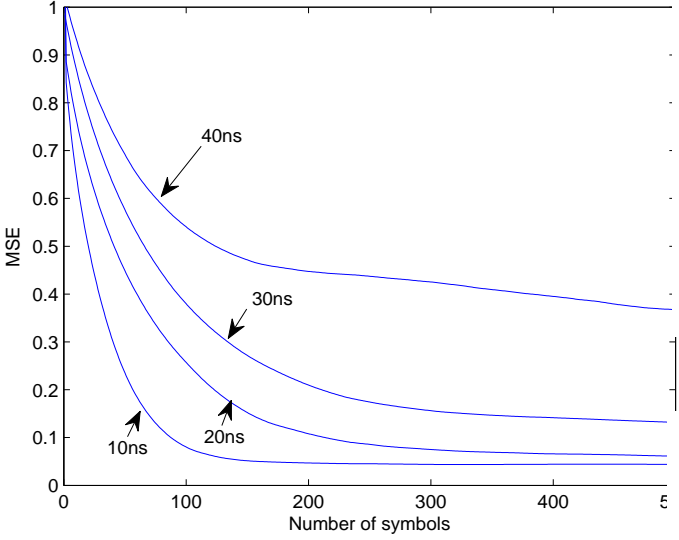,scale=0.7}}
\end{minipage}
\caption{MSE performance of the blind channel estimation (with ChMo2).}
\label{fig:channeldelay}
\end{figure}
%%%%%%%%%%%%%%%%%%%%%%%%%%%%%%%%%%%%%%%%%%%%%%%%%%%%%%%%%%%%%%%%%%%%%
%%%%%%%%%%%%%%%%%%%%%%%%%%%%%%%%%%%%%%%%%%%%%%%%%%%%%%%%%%%%%%%%%%%%%
\begin{figure}[htb]
\begin{minipage}[h]{1.0\linewidth}
  \centering
  \centerline{\epsfig{figure=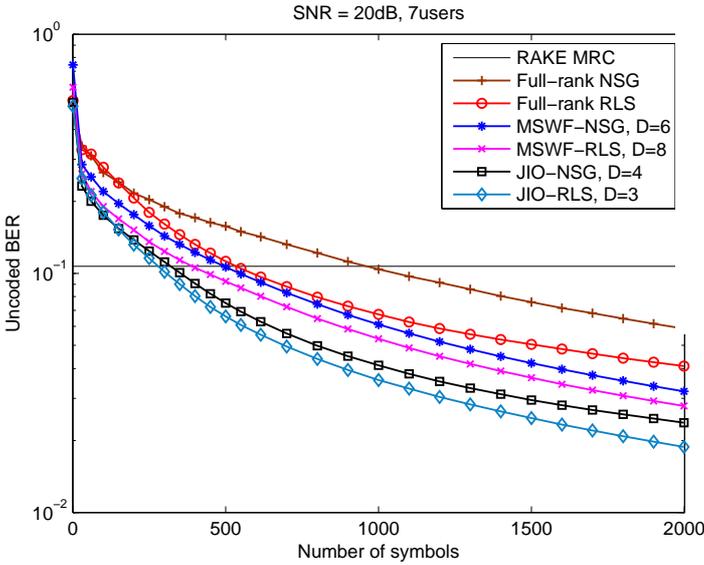,scale=0.7}}
\end{minipage}
\caption{Uncoded BER performance of different algorithms with ChMo2. For full-rank NSG:
$\mu=0.025$, full-rank RLS: $\delta=10$, $\lambda=0.9998$. For MSWF-NSG, $D=6$,
$\mu=0.025$; MSWF-RLS: $D=8$, $\lambda=0.998$. For JIO-NSG $D=4$, $c_{max}=3$,
$v=1$, $\mu_{T,0}=0.075$, $\mu_{w,0}=0.005$; JIO-RLS: $D=3$,
$\lambda=0.9998$, $\delta=10$, $v=0.5$.} \label{fig:bervssymbols}
\end{figure}
%%%%%%%%%%%%%%%%%%%%%%%%%%%%%%%%%%%%%%%%%%%%%%%%%%%%%%%%%%%%%%%%%%%%%
%%%%%%%%%%%%%%%%%%%%%%%%%%%%%%%%%%%%%%%%%%%%%%%%%%%%%%%%%%%%%%%%%%%%%
\begin{figure}[htb]
\begin{minipage}[h]{1.0\linewidth}
  \centering
  \centerline{\epsfig{figure=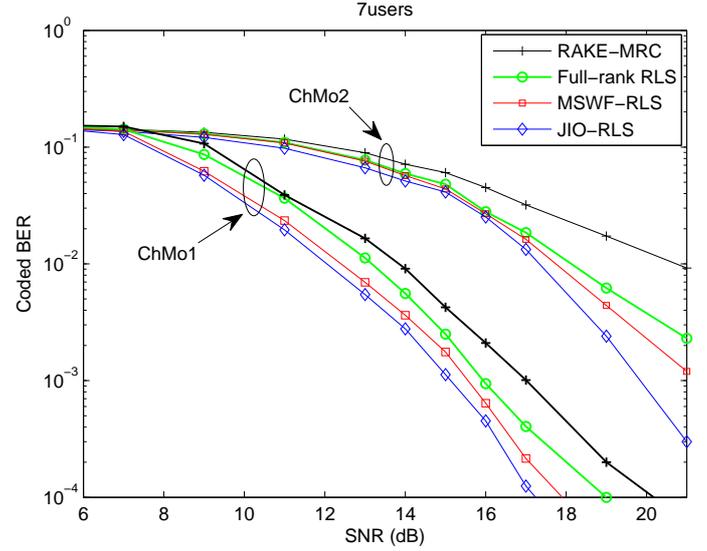,scale=0.7}}
\end{minipage}
\caption{BER performance of the proposed scheme with different SNRs.}
\label{fig:bervssnr}
\end{figure}
%%%%%%%%%%%%%%%%%%%%%%%%%%%%%%%%%%%%%%%%%%%%%%%%%%%%%%%%%%%%%%%%%%%%%
%%%%%%%%%%%%%%%%%%%%%%%%%%%%%%%%%%%%%%%%%%%%%%%%%%%%%%%%%%%%%%%%%%%%%
\begin{figure}[htb]
\begin{minipage}[h]{1.0\linewidth}
  \centering
  \centerline{\epsfig{figure=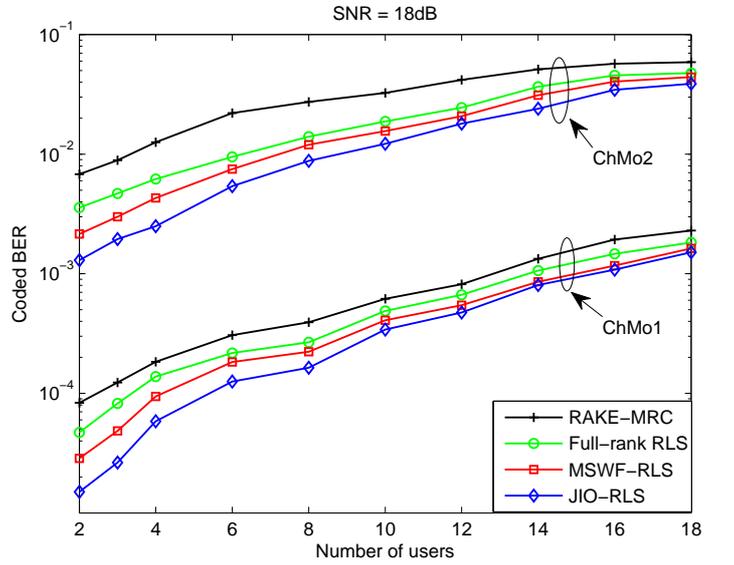,scale=0.7}}
\end{minipage}
\caption{BER performance of the proposed scheme with different number of
users.} \label{fig:bervsuser}
\end{figure}
%%%%%%%%%%%%%%%%%%%%%%%%%%%%%%%%%%%%%%%%%%%%%%%%%%%%%%%%%%%%%%%%%%%%%
%%%%%%%%%%%%%%%%%%%%%%%%%%%%%%%%%%%%%%%%%%%%%%%%%%%%%%%%%%%%%%%%%%%%%
\begin{figure}[htb]
\begin{minipage}[h]{1.0\linewidth}
  \centering
  \centerline{\epsfig{figure=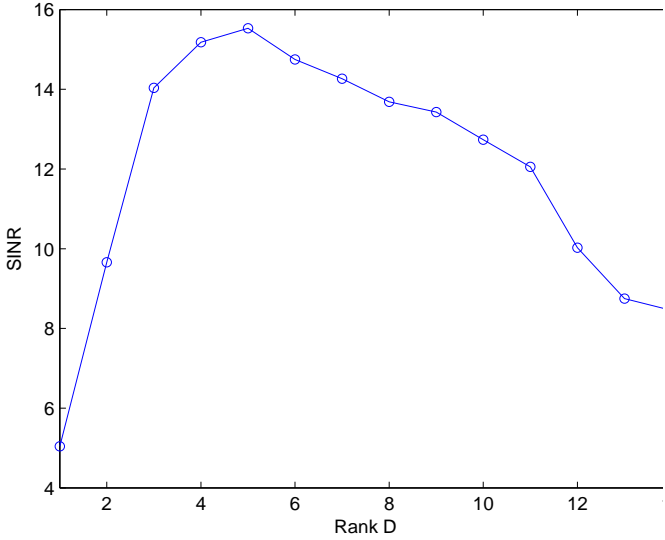,scale=0.7}}
\end{minipage}
\caption{SINR performance with different rank D (with ChMo2).} \label{fig:sinrvsD}
\end{figure}
%%%%%%%%%%%%%%%%%%%%%%%%%%%%%%%%%%%%%%%%%%%%%%%%%%%%%%%%%%%%%%%%%%%%%
%%%%%%%%%%%%%%%%%%%%%%%%%%%%%%%%%%%%%%%%%%%%%%%%%%%%%%%%%%%%%%%%%%%%%
\begin{figure}[htb]
\begin{minipage}[h]{1.0\linewidth}
  \centering
  \centerline{\epsfig{figure=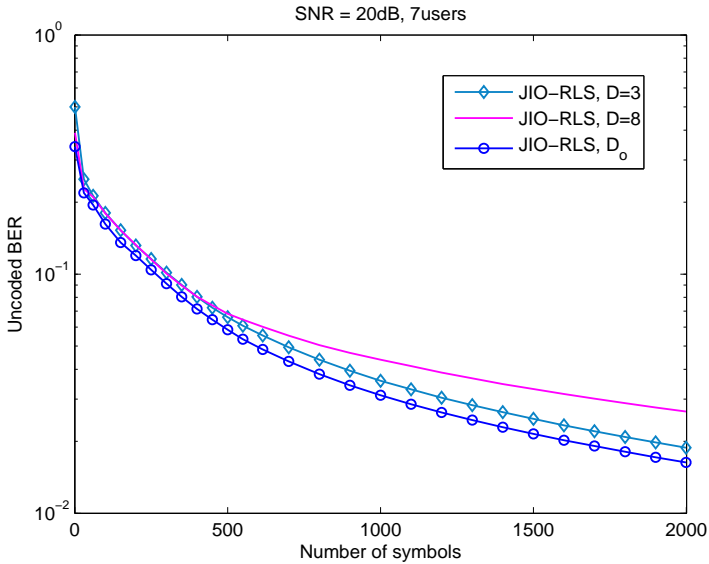,scale=0.7}}
\end{minipage}
\caption{Uncoded BER performance of the rank-adaptation algorithm (with ChMo2).}
\label{fig:dadaptation}
\end{figure}
%%%%%%%%%%%%%%%%%%%%%%%%%%%%%%%%%%%%%%%%%%%%%%%%%%%%%%%%%%%%%%%%%%%%%
%%%%%%%%%%%%%%%%%%%%%%%%%%%%%%%%%%%%%%%%%%%%%%%%%%%%%%%%%%%%%%%%%%%%%%
%\begin{figure}[htb]
%\begin{minipage}[h]{1.0\linewidth}
%  \centering
%  \centerline{\epsfig{figure=berper_coded_blind.eps,scale=0.7}}
%\end{minipage}
%\caption{Coded BER performance with different SNRs.} \label{fig:codebervssnr}
%\end{figure}
%%%%%%%%%%%%%%%%%%%%%%%%%%%%%%%%%%%%%%%%%%%%%%%%%%%%%%%%%%%%%%%%%%%%%%
%%%%%%%%%%%%%%%%%%%%%%%%%%%%%%%%%%%%%%%%%%%%%%%%%%%%%%%%%%%%%%%%%%%%%
\begin{figure}[htb]
\begin{minipage}[h]{1.0\linewidth}
  \centering
  \centerline{\epsfig{figure=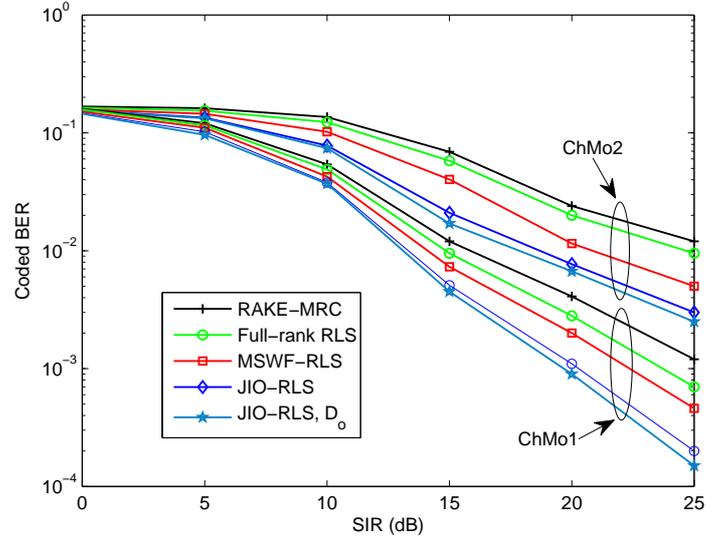,scale=0.7}}
\end{minipage}
\caption{BER performance of the adaptive algorithm with NBI. For NBI,
$f_{d}=23MHz$.} \label{fig:withNBI}
\end{figure}
%%%%%%%%%%%%%%%%%%%%%%%%%%%%%%%%%%%%%%%%%%%%%%%%%%%%%%%%%%%%%%%%%%%%%
{ Firstly, we access the mean squared error (MSE) performance of the
blind channel estimator that is introduced in section
\ref{sec:receiverdesign} with the ChMo2. As shown in
Fig.\ref{fig:channeldelay}, the MSE performance is shown as a
function of number of transmitted symbols with different channel
delays in a 7-user scenario with a signal-to-noise ratio (SNR) of
20dB.} The performance of the blind channel estimation is highly
dependent on the channel delays. The performance of the CCM-based
blind adaptive algorithms will be degraded significantly in the
scenarios of large channel delays due to the inaccuracy of the blind
channel estimation. In this work, we consider a channel delay of
10ns.

{ In Fig.\ref{fig:bervssymbols}, we compare the uncoded BER
performance of the proposed JIO receivers with the full-rank NSG and
RLS algorithms, the MSWF-NSG and MSWF-RLS in the NLOS environment
(ChMo2). In a 7-user scenario with a SNR of 20dB, the uncoded BER
performance of different algorithms as a function of symbols
transmitted is presented that enables us to compare the convergence
rate of different adaptive algorithms. Among all the analyzed
algorithms, the proposed JIO-RLS algorithm converges fastest. The
JIO-NSG algorithm outperforms the MSWF versions and the full-rank
versions with a low complexity. A noticeable improvement on the BER
performance is obtained by using the JIO receivers.}

{ In Fig.\ref{fig:bervssnr} and Fig. \ref{fig:bervsuser}, we access
the coded BER performances of the blind algorithms with different
SNRs in a $7$-user scenario and with different numbers of users in a
18dB SNR scenario, respectively. Both ChMo1 and ChMo2 are considered
in these simulations. The parameters set for all the adaptive
algorithms are the same as in Fig.\ref{fig:bervssymbols}. The
proposed JIO versions show better MAI and ISI canceling capability
in all the simulated scenarios. It can be observed that the use of
channel coding improves the performance of the receivers and that
the same hierarchy in terms of BER performance is verified - the
proposed JIO-RLS algorithm achieves the best performance. In
Fig.\ref{fig:bervssnr}, the JIO-RLS can save around 2dB in
comparison with the MSWF-RLS with ChMo2 for a BER around $10^{-3}$
and save around 1dB with ChMo1 for a BER around $10^{-4}$. In
Fig.\ref{fig:bervsuser}, the JIO-RLS scheme can support more than 2
additional users in comparison with the MSWF-RLS with ChMo2 for a
BER around $10^{-3}$ and can support over 1 additional users with
ChMo1 for a BER around $10^{-4}$.}

{ In Fig.\ref{fig:sinrvsD}, the signal to interference plus noise
ratio (SINR) performance is shown as a function of rank D in the
NLOS environment (ChMo2). We consider a 7-user scenario with a SNR
of 20dB.} A noticeable better performance is obtained for the ranks
in the range of 3 to 8. In this scenario, D = 5 performs best and a
1.5dB gain is achieved compare to the algorithm with D = 3 and D =
8. Note that, for the JIO-RLS algorithm, the complexity is
$O(DM^{2})$. The designer can choose the rank D as a parameter that
will affect the complexity and the performance.

{ Fig.\ref{fig:dadaptation} compares the uncoded BER performance in
the NLOS environment (ChMo2) of the JIO-RLS using the
rank-adaptation algorithm given by \eqref{eq:rankadp} with $D_{\rm
max}=8$ and $D_{\rm min}=3$.} The results using a fixed-rank of 3
and 8 are also shown for comparison purposes and illustration of the
sensitivity of the JIO scheme to the rank D. The forgetting factor
is $\lambda_{D}=0.998$. It can be seen that the uncoded BER
performance of the JIO-RLS scheme with the rank-adaptation algorithm
outperforms the fixed-rank scenarios with $D_{\rm min}=3$ and
$D_{\rm max}=8$. In this experiment, $D=3$ has better steady state
performance than $D=8$, with both cases showing the similar
convergence speed. The rank-adaptation algorithm provides a better
solution than the fixed rank approaches. It should be noted that the
complexity of updating the transformation matrix and the
reduced-rank filter in the rank adaptation algorithm is the same as
the fixed rank case with $D=D_{\rm max}$. Additional complexity is
required to compute the values of $\mathscr{C}_{D}(i)$ by using
\eqref{eq:rankadaptationso} and select the minimum value of a
$(D_{\rm max}-D_{\rm min}+1)$-dimensional vector.

%In Fig.\ref{fig:codebervssnr}, a convolutional code with a
%coding-rate of 2/3 is adopted for the simulations. The code polynomial is
%[7,5,5] and the constraint length is set to 5. In a 7-user scenario, the coded
%BER performance of the RLS-based algorithms is shown as a function of the SNRs.
%Compared to the MSWF-RLS algorithm, the proposed JIO-RLS scheme achieves a
%1.5dB gain for a BER performance of $10^{-3}$. It should be noted that other
%coding schemes employing Turbo codes, LDPC codes and/or iterative detection
%\cite{Rodrigo2008DF} can be employed to further improve the performance of the
%proposed algorithms. Moreover, it can be observed that the use of channel
%coding improves the performance of the receivers and that the same hierarchy in
%terms of BER performance is verified - the proposed JIO-RLS algorithm achieves
%the best performance.

%%%%%%%%%%%%%%%%%%%%%%%%%%%%%%%%%%%%%%%%%%%%%%%%%%%%%%%%%%%%%%%%%%%%%%%%%%%%%%%%%%%%%%%%%%%%%%%%%%%%%%%%
In the last experiment, we examine the blind adaptive algorithms
with an additional narrow band interference (NBI), which is modeled
as a single-tone signal (complex baseband) \cite{xchu2004}:
\begin{equation}
J(t)=\sqrt{P_{j}}e^{\left(2\pi f_{d}t+\theta_{j}\right)},
\end{equation}where $P_{j}$ is the NBI power, $f_{d}$ is the frequency
difference between the carrier frequency of the UWB signal and the one of the
NBI and $\theta_{j}$ is the random phase which is uniformly distributed in
$[0,\pi)$. Here, the received signal can be expressed as
%%%%%%%%%%%%%%%%%%%%%%%%%%%%%%%%%%%%%%%%%%%%%%%%%%%%%%%%%%%%%%%%%%%%%%%%%%%%%%%%%%%%%%%%%%%%%%%%%%%%%%%%
\begin{equation}
z(t)= \sum_{k=1}^{K}\sum_{l=0}^{L-1} h_{k,l}x^{(k)}(t-lT_{\tau})+n(t)+J(t).
\end{equation} Note that, in this experiment, the receivers
are required to suppress the ISI, MAI and NBI together blindly. {In
Fig. \ref{fig:withNBI}, in a 5-user system with a 18dB SNR, the
coded BER performance of the RLS versions are compared with
different signal to NBI ratio (SIR) with ChMo1 and ChMo2.} The
algorithms are set the same parameters as in
Fig.\ref{fig:bervssymbols}. With the NBI, the eigenvalue spread of
the covariance matrix of the received signal is increased and this
change slows down the convergence rate of the full-rank scheme.
However, the proposed JIO receiver shows better ability to cope with
this change and the performance gain over the full-rank scheme is
increased compared to the NBI free scenarios. By adopting the rank
adaptation algorithm, the performance is improved as compared to the
fixed rank JIO-RLS receiver in the simulated scenarios. This is
mainly because of the faster convergence speed that is introduced by
the rank adaptation algorithm.

%%%%%%%%%%%%%%%%%%%%%%%%%%%%%%%%%%%%%%%%%%%%%%%%%%%%%%%%%%%%%%%%%%%%%%%%%%%%%%%%%%%%%%%%%%%%%%%%%%%%%%%%%%%%%%%%%%%
%%%%%%%%%%%%%%%%%%%%%%%%%%%%%%%%%%%%%%%%%%%%%%%%%%%%%%%%%%%%%%%%%%%%%%%%%%%%%%%%%%%%%%%%%%%%%%%%%%%%%%%%%%%%%%%%%%%
\section{Conclusions}
\label{sec:conclusion}

A novel blind reduced-rank receiver is proposed based on JIO and the CCM
criterion. The novel receiver consists of a transformation matrix and a
reduced-rank filter. The NSG and RLS adaptive algorithms are developed for
updating its parameters. In DS-UWB systems, both versions (NSG and RLS) of the
proposed blind reduced-rank receivers outperform the analyzed CCM based
full-rank and existing reduced-rank adaptive schemes with a low complexity. The
robustness of the proposed receivers has been demonstrated in the scenario that
the blind receivers are required to suppress the ISI, MAI and NBI together. The
proposed blind receivers can be employed in spread-spectrum systems which
encounter large filter problems and suffer from severe interferences.

\appendices
%%%%%%%%%%%%%%%%%%%%%%%%%%%%%%%%%%%%%%%%%%%%%%%%%%%%%%%%%%%%%%%%%%%%%%%%%%%%%%%%%%%%%%%%%%%%%%%%%%%%%%%%
\section{Convergence Properties}
\label{app:convergence}

In this section, we examine the convergence properties of the cost function
$\mathbf J_{\rm JIO} =\frac{1}{2}E\left[(\left|y(i)\right|^{2}-1)^{2}\right]$,
where $y(i)=\bar{\mathbf w} ^{H}(i) \mathbf T^{H}(i) \mathbf r(i)$. For
simplicity of the following analysis, we drop the time index $(i)$. The
received signal is given by
%%%%%%%%%%%%%%%%%%%%%%%%%%%%%%%%%%%%%%%%%%%%%%%%%%%%%%%%%%%%%%%%%%%%%%%%%%%%%%%%%%%%%%%%%%%%%%%%%%%%%%%%
\begin{equation}
\begin{split}
\mathbf r&=\sum_{k=1}^{K}\sqrt{E_{k}}\mathbf P_{r}\mathbf
{S}_{e,k}\mathbf h_{k}b_{k}+\mbox{\boldmath$\eta$}+\mathbf
{n},\\
&=\sum_{k=1}^{K}\sqrt{E_{k}}b_{k}\mathbf
p_{k}+\mbox{\boldmath$\eta$}+\mathbf {n}=\mathbf {P}_{k}\mathbf
A_{k}\mathbf b +\mbox{\boldmath$\eta$}+\mathbf {n},
\end{split}
\end{equation}where $\mathbf
p_{k}=\mathbf P_{r}\mathbf {S}_{e,k}\mathbf h_{k}$, $k=1,\dots,K$,
are the signature vectors of the users. $\mathbf {P}_{k}=[\mathbf
p_{1},\dots,\mathbf p_{K}]$, $\mathbf
A_{k}=diag(\sqrt{E_{1}},\dots,\sqrt{E_{K}})$ and $\mathbf
b=[b_{1},\dots,b_{K}]$. $\mbox{\boldmath$\eta$}$ and $\mathbf {n}$
represent the ISI and AWGN, respectively. We assume that $b_{k}$,
$k=1,\dots,K$, are statistically independent i.i.d random variables
with zero mean and unit variance and are independent to the noises.
Firstly, we will discuss the noise-free scenario for the analysis,
in which, the output signal of the JIO receiver is given by
%%%%%%%%%%%%%%%%%%%%%%%%%%%%%%%%%%%%%%%%%%%%%%%%%%%%%%%%%%%%%%%%%%%%%%%%%%%%%%%%%%%%%%%%%%%%%%%%%%%%%%%%
\begin{equation}
y=\bar{\mathbf w}^{H}\mathbf T^{H}\mathbf {P}_{k}\mathbf
A_{k}\mathbf b=\mbox{\boldmath$\epsilon$}^{H}\mathbf b,
\end{equation}where $\mbox{\boldmath$\epsilon$}\triangleq \mathbf
A_{k}^{H}\mathbf {P}_{k}^{H}\mathbf T\bar{\mathbf
w}=[\mbox{$\epsilon$}_{1},\dots,\mbox{$\epsilon$}_{K}]$. Assuming
that user 1 is the desired user and recalling the constraint
$\bar{\mathbf w}^{H}\mathbf T^{H}\mathbf p_{1}=v$, where $v$
is a real-valued constant. We obtain that the first element of the
vector $\mbox{\boldmath$\epsilon$}$ can be expressed as
%%%%%%%%%%%%%%%%%%%%%%%%%%%%%%%%%%%%%%%%%%%%%%%%%%%%%%%%%%%%%%%%%%%%%%%%%%%%%%%%%%%%%%%%%%%%%%%%%%%%%%%%
\begin{equation}
\mbox{$\epsilon$}_{1}=\sqrt{E_{1}}\mathbf p_{1}^{H}\mathbf
T\bar{\mathbf w}=\sqrt{E_{1}}v.
\end{equation}

Now, let us have a closer look at the cost function,
%%%%%%%%%%%%%%%%%%%%%%%%%%%%%%%%%%%%%%%%%%%%%%%%%%%%%%%%%%%%%%%%%%%%%%%%%%%%%%%%%%%%%%%%%%%%%%%%%%%%%%%%
\begin{equation}
\begin{split}
\mathbf J_{\rm JIO}
&=\frac{1}{2}E\left[\left|y(i)\right|^{4}-2\left|y(i)\right|^{2}+1\right]\\
&=\frac{1}{2}\left(E\left[(\mbox{\boldmath$\epsilon$}^{H}\mathbf b\mathbf
b^{H}\mbox{\boldmath$\epsilon$})^{2}\right]-2E\left[\mbox{\boldmath$\epsilon$}^{H}\mathbf
b\mathbf b^{H}\mbox{\boldmath$\epsilon$}\right]+1\right)\\
&=\frac{1}{2}\left(\sum^{K}_{k=1}\sum^{K}_{j=1}|\mbox{$\epsilon$}_{k}|^{2}|\mbox{$\epsilon$}_{j}|^{2}|b_{k}|^{2}|b_{j}|^{2}
-2\sum^{K}_{k=1}|\mbox{$\epsilon$}_{k}|^{2}|b_{k}|^{2}+1\right)\\
&=\frac{1}{2}\left(\sum^{K}_{k=1}\sum^{K}_{j=1}|\mbox{$\epsilon$}_{k}|^{2}|\mbox{$\epsilon$}_{j}|^{2}
-2\sum^{K}_{k=1}|\mbox{$\epsilon$}_{k}|^{2}+1\right)\\
&=\frac{1}{2}(|\mbox{$\epsilon$}_{1}|^{2}+\tilde{\mbox{\boldmath$\epsilon$}}^{H}\tilde{\mbox{\boldmath$\epsilon$}})^{2}
-(|\mbox{$\epsilon$}_{1}|^{2}+\tilde{\mbox{\boldmath$\epsilon$}}^{H}\tilde{\mbox{\boldmath$\epsilon$}})+\frac{1}{2}\\
\end{split}
\label{eq:mainanalysis}
\end{equation}where $\tilde{\mbox{\boldmath$\epsilon$}}=[\mbox{$\epsilon$}_{2},\dots,\mbox{$\epsilon$}_{k}]
= \tilde{\mathbf A}_{k}^{H}\tilde{\mathbf {P}}_{k}^{H}\mathbf
T\bar{\mathbf w}$, $\tilde{\mathbf {P}}_{k}=[\mathbf
p_{2},\dots,\mathbf p_{K}]$ and $\tilde{\mathbf
A}_{k}=diag(\sqrt{E_{2}},\dots,\sqrt{E_{K}})$. Equation
\eqref{eq:mainanalysis} transforms the cost function of both
$\mathbf T$ and $\bar{\mathbf w}$ into a function with single
variable $\tilde{\mbox{\boldmath$\epsilon$}}$. We remark that
$\tilde{\mbox{\boldmath$\epsilon$}}$ is a linear function of
$\mathbf T\bar{\mathbf w}$ that is the blind reduced-rank receiver.
Hence, the convexity properties of the cost function with respect to
$\tilde{\mbox{\boldmath$\epsilon$}}$ reflects the convexity
properties of the cost function with respect to $\mathbf
T\bar{\mathbf w}$. To evaluate the convexity of $\mathbf J_{\rm
JIO}$, we compute its Hessian that is given by
%%%%%%%%%%%%%%%%%%%%%%%%%%%%%%%%%%%%%%%%%%%%%%%%%%%%%%%%%%%%%%%%%%%%%%%%%%%%%%%%%%%%%%%%%%%%%%%%%%%%%%%%
\begin{equation}
\mathbf H_{JIO}=\frac{\partial}{\partial
\tilde{\mbox{\boldmath$\epsilon$}}^{H}}\frac{\partial\mathbf J_{\rm
JIO}}{\partial
\tilde{\mbox{\boldmath$\epsilon$}}}=2\tilde{\mbox{\boldmath$\epsilon$}}\tilde{\mbox{\boldmath$\epsilon$}}^{H}
+(|\mbox{$\epsilon$}_{1}|^{2}-1)\mathbf I.
\end{equation}It can be concluded that a sufficient condition for $\mathbf H_{JIO}$ to be a positive definite
matrix is $|\mbox{$\epsilon$}_{1}|^{2}>1$, which is $E_{1}v^{2}>1$. This
condition is obtained in noiseless scenario, however, it also holds for small
$\sigma^{2}$ that can be considered as a slight perturbation of the noise-free
case \cite{CXU2001}. For larger values of $\sigma^{2}$, the term $v$ can be
adjusted to ensure the convexity of the cost function.

%%%%%%%%%%%%%%%%%%%%%%%%%%%%%%%%%%%%%%%%%%%%%%%%%%%%%%%%%%%%%%%%%%%%%%%%%%%%%%%%%%%%%%%%%%%%%%%%%%%%%%%%

\end{document}